\documentclass[prb,twocolumn,showpacs,superscriptaddress,floatfix,aps,longbibliography]{revtex4-1}

\usepackage[titletoc,toc,title]{appendix}
\usepackage{amsmath,amssymb,mathtools,verbatim}
\usepackage{graphicx}
\usepackage[font=small]{caption}
\usepackage{subcaption}
\usepackage{color}
\usepackage{xcolor}
\usepackage{rotating}
\usepackage{bm}
\usepackage{setspace}
\usepackage{hyperref}
\usepackage{float}
\usepackage{soul}
\usepackage{color}
\usepackage{braket}
\usepackage{physics}
\usepackage{bbold}
\usepackage[export]{adjustbox}
\usepackage{booktabs}
\usepackage{physics}
\graphicspath{{figures/}}

\newcommand{\misfitr}[3]{ V^{\mathrm{GSFE}}_{#1}(\bm{B}^{#2 \rightarrow #3} (\omega))}

\newcommand{\misfit}[3]{ \Phi^{\mathrm{misfit}}_{#1}(\bm{b}^{#2 \rightarrow #3} (\omega))}

\newcommand{\inter} {\Phi^{\mathrm{inter}}_{i, \omega} (\bm{u})}
\newcommand{\intra} {\Phi^{\mathrm{intra}}_{i, \omega} (\bm{u})}
\newcommand{\rr}[1]{\mathrm{#1}}
\newcommand{\bu}{\bm{u}}
\newcommand{\gradx}{\grad_{ \bm{r}}}
\newcommand{\e}{\operatorname{e}}

\newcommand{\bij}{\bm{b}^{i \rightarrow j}}
\newcommand{\Bij}{\bm{B}^{i\rightarrow j}}

\hypersetup{
colorlinks=true,
urlcolor= blue,
citecolor=blue,
linkcolor= blue,
bookmarks=true,
bookmarksopen=false,
}

\renewcommand{\vec}[1]{\bm{#1}}

\begin{document}

\preprint{APS/123-QED}

\title{Moir\'e of Moir\'e: Modeling Mechanical Relaxation in Incommensurate Trilayer van der Waals Heterostructures}

\author{Ziyan Zhu}
\affiliation{Department of Physics, Harvard University, Cambridge, Massachusetts 02138, USA}
\author{Paul Cazeaux}
\affiliation{Department of Mathematics, University of Kansas, Lawrence, Kansas 66045, USA}
\author{Mitchell Luskin}
\affiliation{School of Mathematics, University of Minnesota, Minneapolis, Minnesota 55455, USA}
\author{Efthimios Kaxiras}
\affiliation{Department of Physics, Harvard University, Cambridge, Massachusetts 02138, USA}
\affiliation{John A. Paulson School of Engineering and Applied Sciences, Harvard University, Cambridge, Massachusetts 02138, USA}

\date{\today}

\begin{abstract}
The incommensurate stacking of multi-layered two-dimensional materials is a challenging problem from a theoretical perspective and an intriguing avenue for manipulating their physical properties. Here we present a multi-scale model to obtain the mechanical relaxation pattern of twisted trilayer van der Waals (vdW) heterostructures with two independent twist angles, a generally incommensurate system without a supercell description. We adopt the configuration space as a natural description of such incommensurate layered materials, based on the local environment of atomic positions, bypassing the need for commensurate approximations. To obtain the relaxation pattern, we perform energy minimization with respect to the relaxation displacement vectors. We use a continuum model in combination with the Generalized Stacking Fault energy to describe the interlayer coupling, obtained from first-principles calculations based on Density Functional Theory. We show that the relaxation patterns of twisted trilayer graphene and $\mathrm{WSe_2}$ are ``moir\'e of moir\'e", as a result of the incommensurate coupling two bilayer moir\'e patterns. We also show that, in contrast to the symmetry-preserving in-plane relaxation in twisted bilayers, trilayer relaxation can break the two-fold rotational symmetry about the xy-plane when the two twist angles are equal.
\end{abstract}

\maketitle

\section{Introduction}
Assemblies of multilayers of two-dimensional (2D) materials, referred to as van der Waals (vdW) heterostructures, have become a favorite platform for exploring strongly correlated states, because they offer a large parameter space and great tunability. Superconductivity and strongly-correlated states have been discovered in these heterostructures with one small twist angle (e.g., twisted bilayer graphene, twisted double-bilayer graphene)~\cite{cao2018unconventional, cao2018correlated, yankowitz2018dynamic, liu2019spin, cao2019electric, william2019correlated, cheng2019observation,lu2019superconductors} and systems with a small misalignment (e.g., hBN + graphene)~\cite{chen2018evidence, moriyama2019observation}. Due to the twist angle or the small lattice misalignment, the system forms moir\'e patterns and the periodicity becomes much larger than the original unit cell size.
To understand the electronic and mechanical properties of the multilayered structures, we view them as a series of conventional crystals with a weak perturbative interaction between layers~\cite{tritsaris2016perturbation}.

Crystal relaxation in vdW heterostructures has been studied in the continuum limit by energy minimization~\cite{san2014spontaneous, dai2016twisted, nam2017lattice, malena2017discrete, malena2018discrete, carr2018relaxation, leconte2019relaxation}. Relaxation in incommensurate stacked 2D materials has been known to form domain walls separating large commensurate areas~\cite{coppersmith1982dislocations}. For example, in twisted bilayer graphene, relaxation enlarges the AB/BA stacking regions (the equilibrium stacking) and forms a thin domain line in between adjacent AB/BA stacking regions, while the AA stacking regions (high energy stacking order) shrink to localized spots. The domain formation in twisted bilayer systems has been observed experimentally~\cite{alden2013strain, woods2014commensurate, van2015relaxation, yankowitz2016pressure,  kim2017evidence, yoo2019atomic}. The relaxation pattern has a dependence on the twist angle: the magnitude of relaxation displacement vectors decreases as the twist angle increases. At a twist angle $\theta > \theta^*$ ($\theta^* \sim 1^\circ$ in twisted bilayer graphene), the relaxation effects are weak. As the twist angle becomes smaller, the domain wall thickness stays roughly constant and only the AB/BA stacking regions get enlarged.

A natural extension is to add another layer, which introduces a new degree of freedom --- a second twist angle --- to tune the system. We emphasize that the system we discuss here is different than multi-layered systems such as double-bilayer graphene, in which there is only one independent parameter or one twist angle despite the presence of four layers. The experimental technique to realize twisted trilayer graphene is readily available and unconventional correlated states have been observed~\cite{tsai2019correlated}. There have been preliminary studies of electronic states of twisted trilayer graphene using $k\cdot p $ perturbation theory~\cite{mora2019flat, ma2019topological,tsai2019correlated}. In these models, mechanical relaxation has not been taken into consideration. However, atomic relaxation will modify the band structure by opening up a single-particle gap in twisted bilayer graphene~\cite{nam2017lattice, carr2019minimal, guinea2019continuum} and thus is an important effect in obtaining realistic description of electronic behaviors.

The extension from twisted bilayers to twisted trilayers is challenging for the following reasons. (1) The system size becomes much larger. In twisted bilayer systems the relevant length scale is on the order of $1/\theta$. In twisted trilayers, assuming the two twist angles are identical, the length scale becomes $1/\theta^2$, where $\theta$ is the relevant twist angle, typically a small quantity for systems of interest. (2) Because of the independent second twist angle, a supercell description is no longer valid, except for specific cases. The trilayer system is truly incommensurate, while there always exists a moir\'e supercell in the continuum limit or the small angle limit of twisted bilayers.

To address the challenge of the large system size and the lack of periodicity, we introduce a multiscale method based on the concept of configuration space to minimize the energy directly in the large system limit without having to use a supercell approximation. Specifically,  we perform energy minimization in the continuum approximation over a collection of local atomic environments with respect to the other layers, which we call the configuration space~\cite{massatt2017electronic, cazeaux2017analysis,cances2017generalized,cazeaux2018energy}. This approach has recently been applied to describing the crystal relaxation in stacked bilayer materials ~\cite{carr2018relaxation,cazeaux2018energy}. 

The rest of the paper is organized as follows: in Section~\ref{sec:config}, we introduce the configuration space to describe the trilayer system and formulate the energy minimization problem in configuration space using a continuum approximation. In Section \ref{sec:results}, we present the results for twisted trilayer graphene and $\mathrm{WSe_2}$, as representative heterostructures of a 2D semi-metal and a 2D semiconductor. Finally, we summarize the results and discuss potential applications in Section \ref{sec:conclusion}.


\begin{figure}[ht!]
\centering
\includegraphics[width=\linewidth, valign=c]{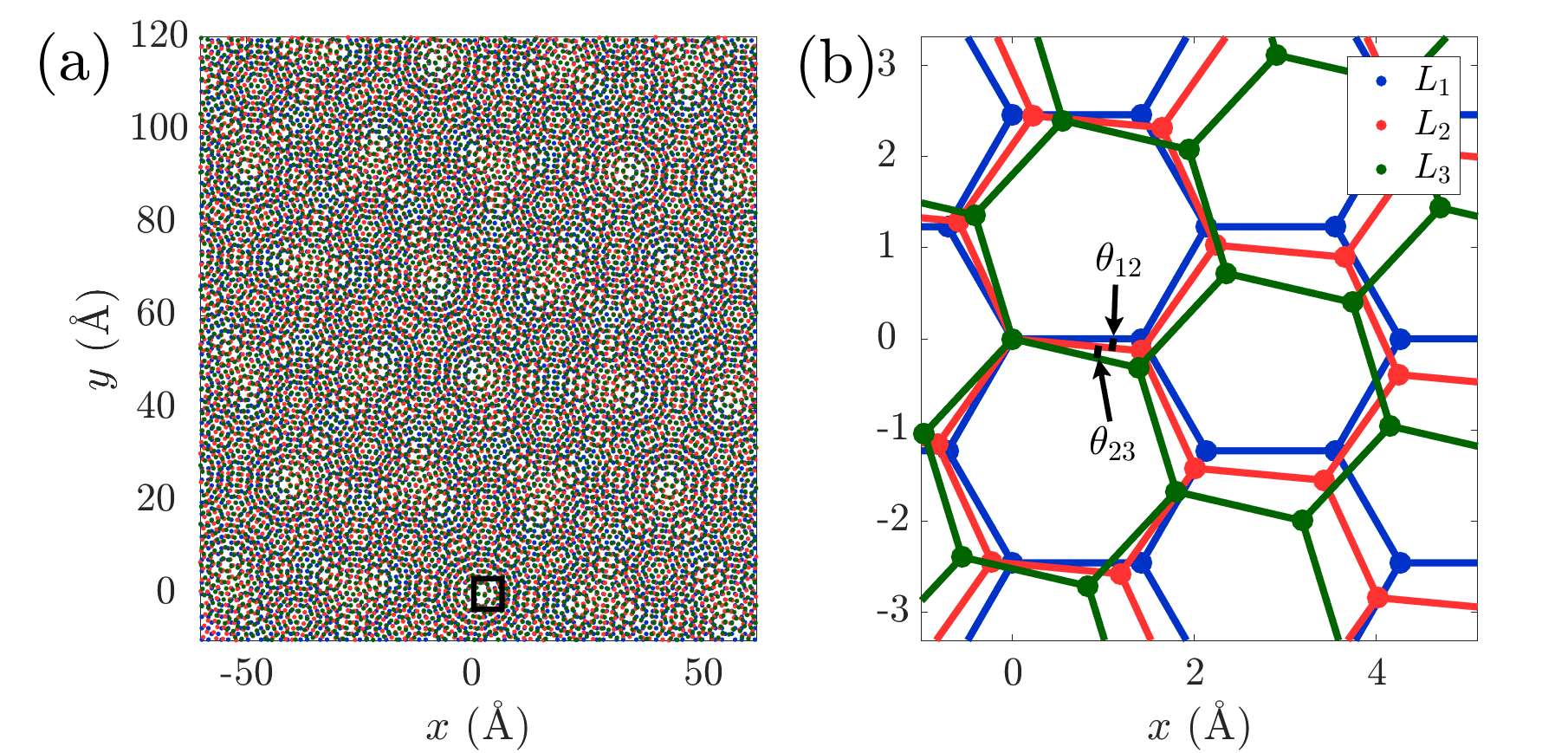}
\caption{(a) An example of a twisted trilayer honeycomb lattice in real space with $\theta_{12} = 5.3^\circ$ and $\theta_{23} = 7.7^\circ$. (b) Magnified view at the black box marked in (a). The twist angle between $L_1$ and $L_2$, $\theta_{12}$, and the twist angle between $L_2$ and $L_3$, $\theta_{23},$ are marked by black arrows.}
\label{fig:geom}
\end{figure}

\section{Continuum Approximation and Energy Minimization in Configuration Space} \label{sec:config}
\subsection{Setup and Notation}
We first define the Bravais lattice basis vectors of a monolayer, $\bm{a}_1$ and $\bm{a}_2$ respectively, in a $2 \times 2$ matrix form as
the columns of the following matrix:
\begin{align} A = a_0
	\begin{pmatrix}
		\sqrt{3}/2 & \sqrt{3}/2 \\ -1/2& 1/2
	\end{pmatrix} = (\bm{a}_1\quad  \bm{a}_2),
\end{align}
where $a_0$ is the lattice constant. We denote the unit cell of layer $i$ by $\Gamma^{(i)}.$

We consider the twisted trilayer systems with two independent twist angles; an example of such a system is given in Fig.~\ref{fig:geom}a. Suppose the second layer $L_2$ is fixed at the origin. The first layer ($L_1$) is twisted counterclockwise by $\theta_{12}$ with respect to $L_2$, and $L_3$ is twisted clockwise by $\theta_{23}$ with respect to $L_2$ (see Fig.~\ref{fig:geom}b). The primitive Bravais lattice basis vectors in the matrix form of $L_i$ is defined as $A_i$. With $A_2 = A$, $A_1$ and $A_3$ are given as follows,
\begin{align}
	A_1 =& \begin{pmatrix}
		\cos{\theta_{12}} & -\sin{\theta_{12}} \\ \sin{\theta_{12}} & \cos{\theta_{12}}
	\end{pmatrix} A
	= (\bm{a}_1^{(1)} \quad \bm{a}_2^{(1)}),\nonumber \\
	A_3 =&
	\begin{pmatrix}
		\cos{\theta_{23}} & \sin{\theta_{23}} \\ -\sin{\theta_{23}} & \cos{\theta_{23}}
	\end{pmatrix} A =  (\bm{a}_1^{(3)} \quad \bm{a}_2^{(3)}).
\end{align}
The primitive reciprocal lattice vectors are given by the columns of $2\pi A_i^{-T}.$

\begin{figure}[ht!]
\centering
	\includegraphics[width=0.8\linewidth, valign=c]{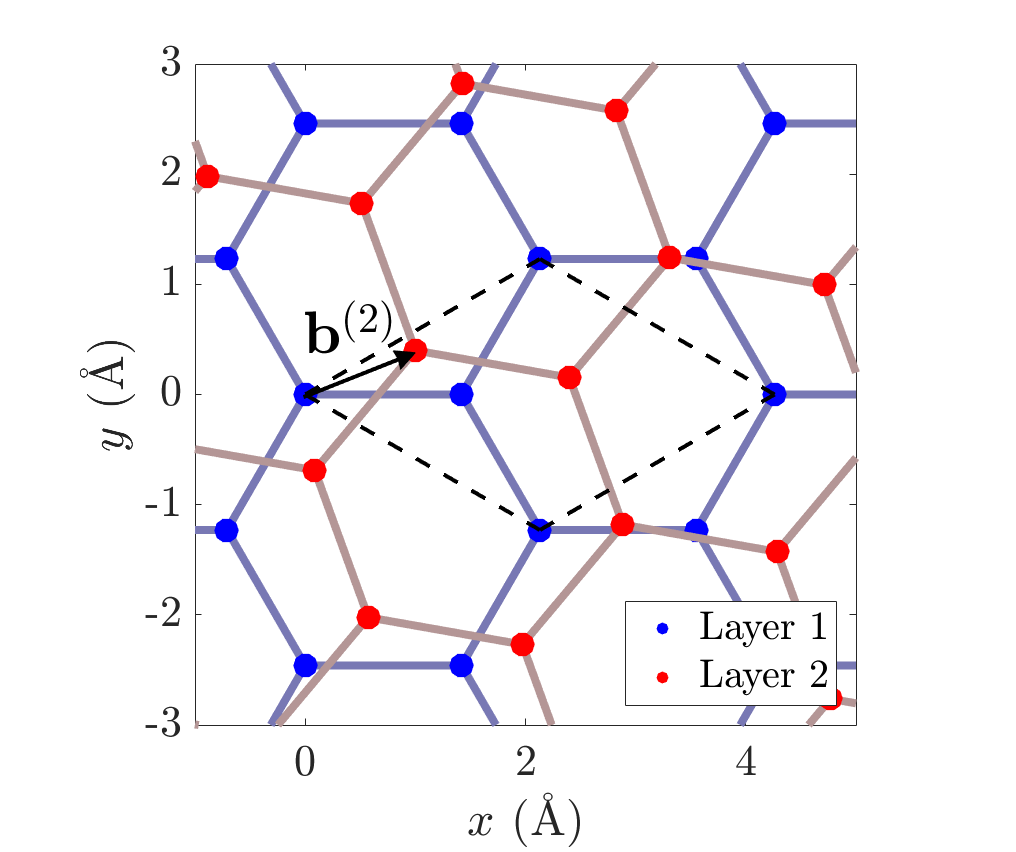}
	\caption{The definition of the local shift or the disregistry $\bm{b}^{(j)}$. Here, we take $j=2$ as an example, defining $\bm{b}^{(2)}$ between $L_1$ and $L_2$.}
	\label{fig:disregistry}
\end{figure}

\begin{figure*}[ht!]
	\centering
	\begin{subfigure}{.505\linewidth}
	\centering
	\includegraphics[width=\textwidth, valign=c]{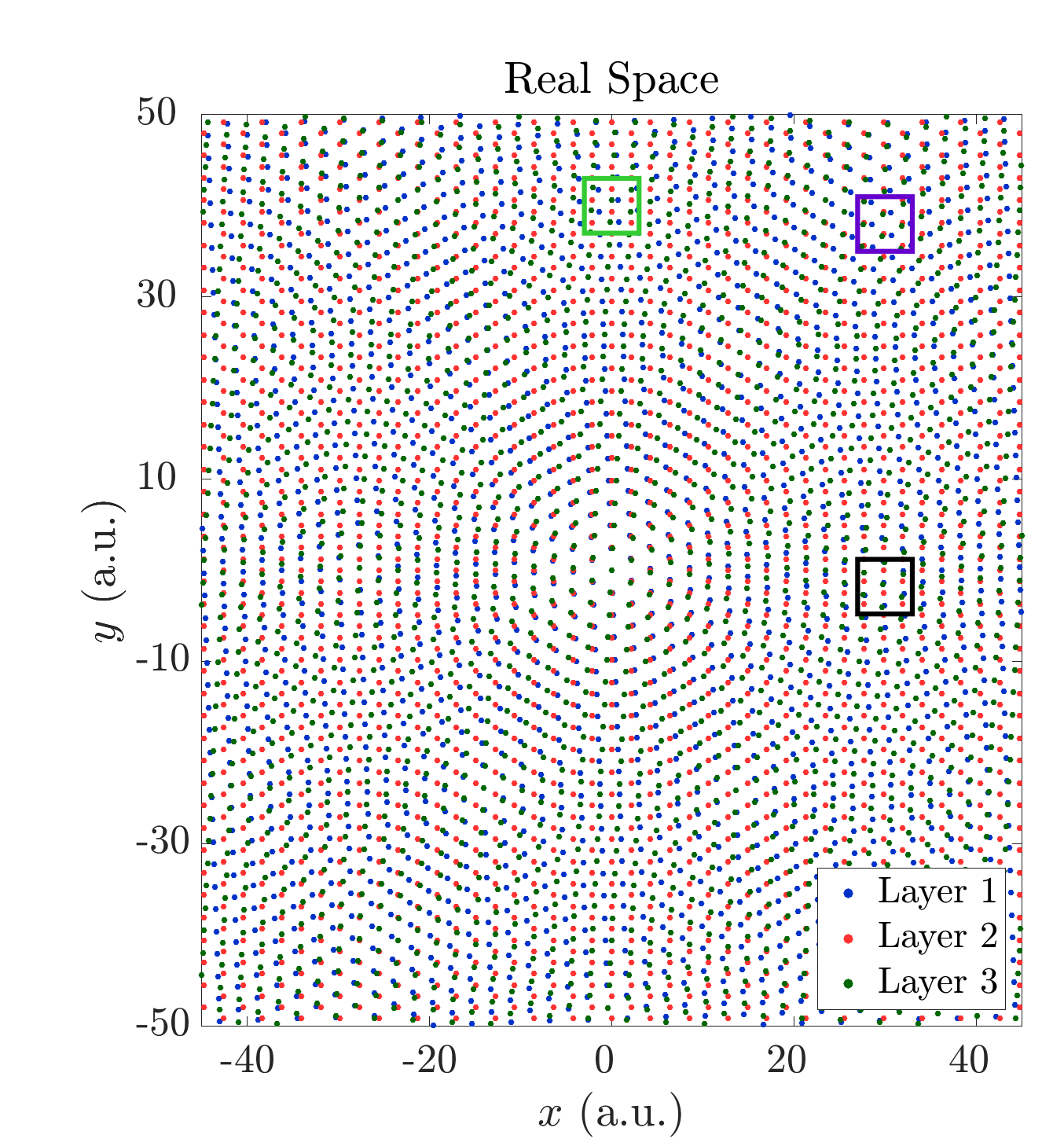}
	\end{subfigure}
	\begin{subfigure}{.2056\linewidth}
	\centering
	\includegraphics[width=\textwidth, valign=c]{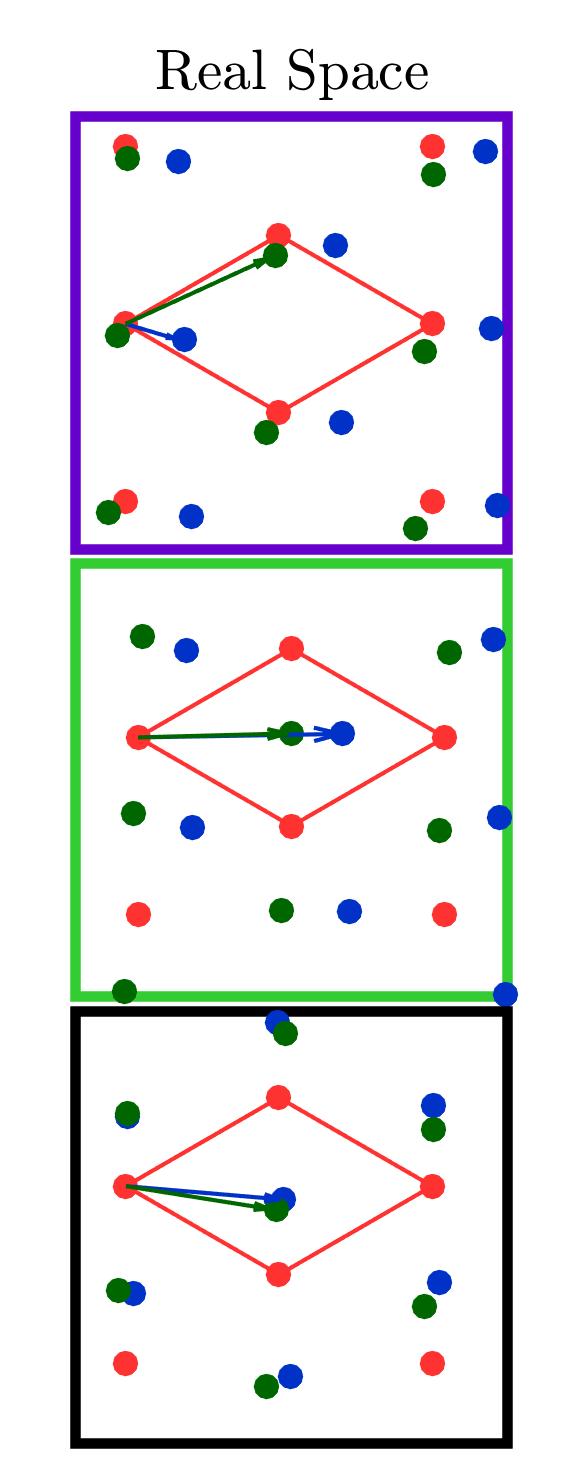}
	\end{subfigure}%
	\begin{subfigure}{.2056\linewidth}
	\centering
	\includegraphics[width=\textwidth,  valign=c]{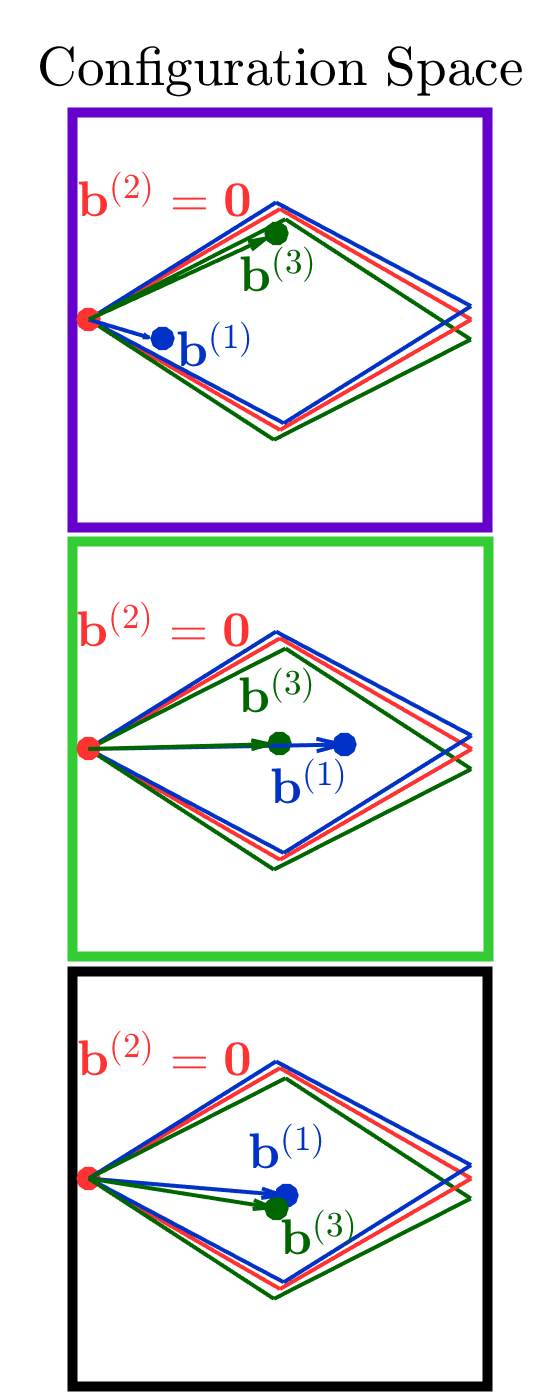}
	\end{subfigure}%
	\caption{Mapping between real and configuration space. Left: Real space atomic positions of a trilayer triangular lattice. The first layer (blue) is twisted by $2^\circ$ with respect to the second layer (red), and the third layer (green) is twisted by $3^\circ$ with respect to the second. Middle: magnified view of the three boxed area on the left. Red parallelograms are the unit cell of the layer 2. Arrows are pointing from the atom of interest (the red atom on the left corner) to the neighboring atoms on the other two layers. Right: the corresponding configuration space to the atom of interest. Blue, red, and green parallelograms are the unit cells of $L_i$ for $i=1,2,3$ respectively.}
\label{fig:setup}
\end{figure*}

\begin{figure}[ht!]
\centering
	 \includegraphics[width=\linewidth, valign=c]{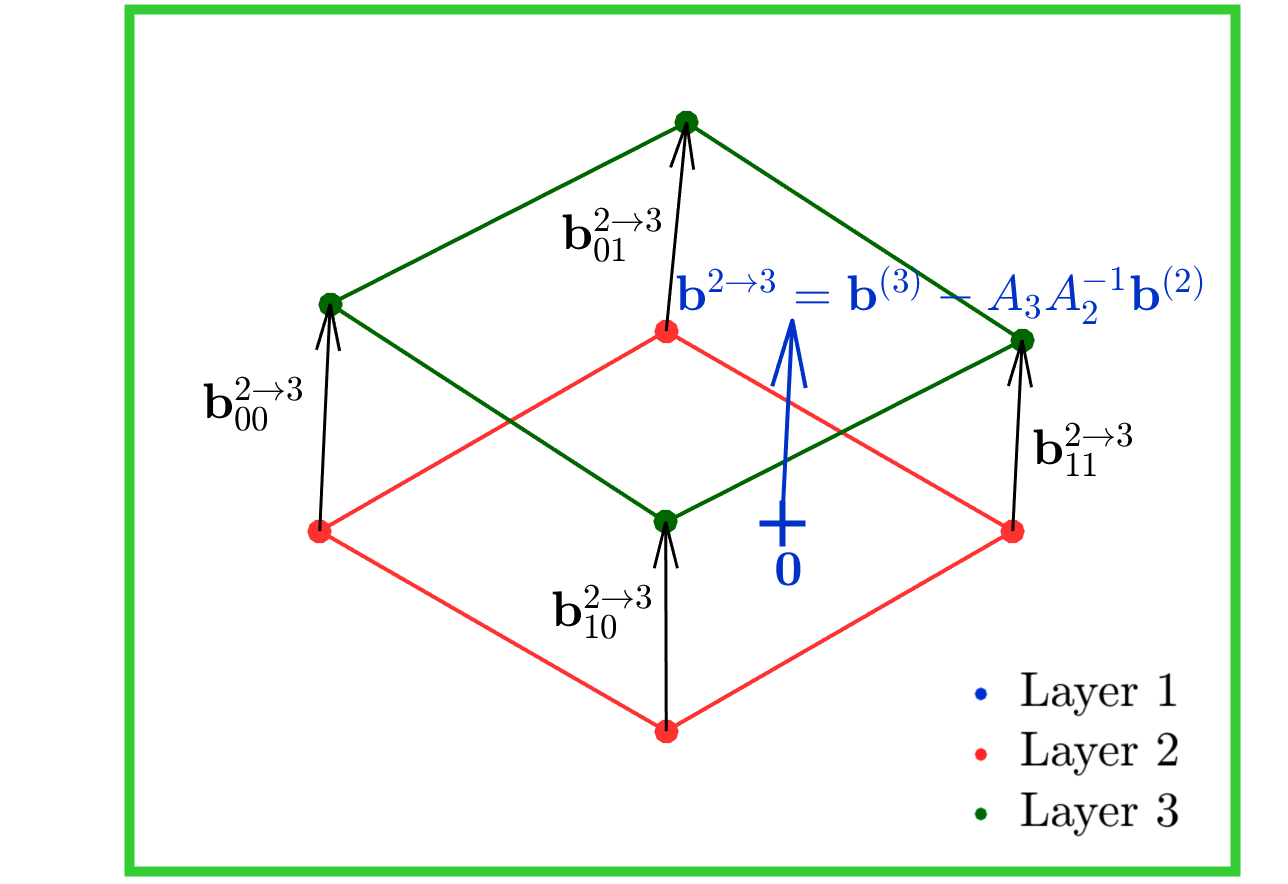}
	\caption{A demonstration of the bilinear interpolation of the local shift vector $\bm{b}^{2\rightarrow 3}$.  This corresponds to the same real space position in the green (the second) box in Fig.~\ref{fig:setup} but with a different origin. Here, the origin is chosen to be a lattice site of $L_1$ with $\bm{b}^{(1)} = \bm{0}$, which is marked by a blue cross in the figure, while in Fig.~\ref{fig:setup}, the origin is the red atom on the left corner. $\bm{b}_{mn}^{2\rightarrow3}$ are the relative shifts between $L_2$ and $L_3$ at the four nearest lattice sites to the origin. }
	\label{fig:interp}
\end{figure}

\subsection{Relaxation in Configuration Space}
The issue with applying the real space continuum model to obtain the relaxation pattern in trilayer systems is the lack of periodicity. While bilayer systems always have a
moir\'e supercell \cite{carr2018relaxation,cazeaux2018energy}, trilayer systems are generally incommensurate and thus lack a periodic supercell even in the continuum limit.
Therefore, a more general description beyond the supercell approximation is required. Here, we introduce the configuration space to describe the local environment of every position in the continuum and parametrize the trilayer system in configuration space~\cite{massatt2017electronic, cazeaux2017analysis,cances2017generalized} based on the formalism introduced by Cazeaux {\it et al.} \cite{cazeaux2018energy}. We also reformulate the energy minimization problem in configuration space.

The local configuration or the environment for an arbitrary position $\bm{r}$ in layer $L_i$, which may be a lattice position on layer $L_i$,
can be uniquely determined by three
relative shift vectors or disregistries $\bm{b}^{(j)}$
describing the relative position from this position in $L_i$ with respect to the lattice of layer $L_j$ (see Fig.~\ref{fig:disregistry} for an example), which is defined as the lattice position with respect to $\bm{r}$ modulo the unit cell $\Gamma^{(j)}$.
More explicitly,
\begin{equation} \label{eqn:bj}
\bm{b}^{(j)} (\bm r) = \bm{R}^{(j)} - \bm{r} \in \Gamma^{(j)},
\end{equation}
where $\bm{R}^{(j)}$ is any lattice site in layer $L_j$.
The shift vectors $\bm{b}^{(j)}$ take values in the periodic unit cell $\Gamma^{(j)}$ of layer $L_j$ by the translational invariance of the lattice.
Fig.~\ref{fig:setup} illustrates this mapping between real space and the configuration space of the three relative shift vectors for $\bm{r}$ sampled at lattice points of the
 middle layer $L_2$ of a triangular lattice trilayer for a representative atomic structure. The collection of these shift vectors forms a six-dimensional configuration space:
 \begin{equation}
 	\Omega = \{\omega \mid \omega = (\bm{b}^{(1)}, \bm{b}^{(2)}, \bm{b}^{(3)})\}.\end{equation}

The shifts $\bm{b}^{(j)}(\bm{r})$ vary quickly on the scale of the lattice size, and thus do not correspond to the intuitive notion of disregistry which varies on the scale of the moir\'e pattern.
Instead, we construct functions $\bij(\omega)$ measuring smoothly the disregistry between layer $L_i$ and $L_j$ as a function of the local configuration $\omega$, by using bilinear interpolation between lattice points~\cite{carr2018relaxation,cazeaux2018energy}
\begin{equation}\label{eqn:bji}
	\bm{b}^{i\rightarrow j} (\omega) = \bm{b}^{(j)} - A_j A_i^{-1} \bm{b}^{(i)} \in \Gamma^{(j)}.
\end{equation}
This expression is periodic as a function of $\bm{b}^{(i)}$ and $\bm{b}^{(j)}$ respectively in the unit cells of $L_i$ and $L_j$, and the disregistry $\bm{b}^{i \rightarrow j}(\omega(\bm{r}))$ varies slowly on the atomic scale.

Fig.~\ref{fig:interp} demonstrates this construction.
We first choose a reference point $\bm{r}$; here, we use a lattice site of $L_1$ as an example. At the reference point, the configuration is $\omega = (\bm{b}^{(1)}, \bm{b}^{(2)}, \bm{b}^{(3)})$.
At each one of the four lattice sites in layer $L_i$ (respectively $L_j$) denoted by $\vec{R}_{mn}^{(i)}$ (respectively $\vec{R}_{mn}^{(j)}$) neighboring the reference point $\bm{r}$, the shift vector $\vec{b}^{(j)}(\vec{R}_{mn}^{(i)})$ coincides with the local disregistry of layer $L_j$ with respect to layer $L_i$:
\begin{equation}\label{eqn:bij_site}
\bm{b}^{i \rightarrow j}_{mn}  =  \bm{R}_{mn}^{(j)} - \bm{R}_{mn}^{(i)} \equiv \bm{b}^{(j)} (\bm{R}_{mn}^{(i)}),
\end{equation}
where we use the definition in Eq.~(\ref{eqn:bj}).
Because the lattices are almost aligned, these four disregistries only differ slightly from each other.
We can then define $\bm{b}^{i \rightarrow j} (\omega)$ by bilinear interpolation of the four relative shifts $\bm{b}_{mn}^{2 \rightarrow 3}$, yielding Eq.~\eqref{eqn:bji}~\cite{cazeaux2018energy}.
Note that the interpolated disregistry given in Eq.~(\ref{eqn:bji}) is consistent with Eq.~(\ref{eqn:bij_site}) at any lattice site of $L_i$.

In practice, we need to only consider reference points that coincide with lattice points in layer $L_i$, $\bm{r} = \bm{R}^{(i)}$, in order to parameterize atomic displacements. This leads us to introduce the reduced, four-dimensional atomic configuration spaces:
\begin{equation}
 	\Omega_i = \{\omega \mid \omega = (\bm{b}^{(1)}, \bm{b}^{(2)}, \bm{b}^{(3)}) \ \vert \ \bm{b}^{(i)} = \bm{0}\}.
\end{equation}
The relaxation displacement $\bu^{(i)}(\omega)$ at each lattice point is a function of its local environment of shifts $\omega\in \Omega_i.$ The above construction allows us to define an interpolated displacement $\widehat \bu^{(j)} (\omega)$ of another layer $L_j$ from the viewpoint 
of a lattice site in $L_i$ 
by
\begin{align} \label{eq:hatu}
	\widehat{\bu}^{(j)} (\omega) 
		= \bu^{(j)} (\bm{b}^{j \rightarrow 1} (\omega), \bm{b}^{j \rightarrow 2} (\omega), \bm{b}^{j \rightarrow 3} (\omega)), 
\end{align}
where we simply use the definition given in Eq.~(\ref{eqn:bji}).
For example, the interpolated displacement $\widehat \bu^{(2)} (\omega)$ for $\omega\in \Omega_1$ is
\begin{align}
	\widehat{\bu}^{(2)} (\omega)
	&= \bu^{(2)} (\bm{b}^{2 \rightarrow 1} (\omega), \bm{b}^{2 \rightarrow 3} (\omega))  \\
	&= \bu^{(2)} (- A_1 A_2^{-1} \bm{b}^{(2)}, \bm{b}^{(3)} - A_3 A_2^{-1} \bm{b}^{(2)}),\nonumber
\end{align}
where we omit $\bm{b}^{j \rightarrow j}(\omega)=0$ from the argument of $\bu^{(j)}.$
We can then describe the relaxation-modulated local shift $\Bij(\omega)$ of $L_i$ with respect to $L_j$ by
\begin{equation}
	\bm{B}^{i\rightarrow j} (\omega) =\bm{b}^{i\rightarrow j}(\omega) + \widehat{\bm{u}}^{(j)}(\omega) - \bm{u}^{(i)} (\omega) \ \mathrm{for}\ \omega \in \Omega_i,
	\label{eqn:local_shift}
\end{equation}
where $\widehat{\bm{u}}^{(j)}(\omega)$ is the interpolated displacement of $L_j.$ 

The relaxed site energy has two contributions, interlayer and intralayer site energies, as a function of the relaxation displacement vectors, $\bm{u} = (\bm{u}^{(1)}, \bm{u}^{(2)}, \bm{u}^{(3)})$. For a given layer $L_i$,
\begin{equation}
	\Phi_{i, \omega} (\bm{u}) = \intra + \inter.
\end{equation}
By the uniform sampling of  the continuum of local configurations due to the ergodicity of incommensurate lattices \cite{cances2017generalized,massatt2017electronic}, we can derive the total energy by integrating the site energies over all configurations \cite{cazeaux2018energy}
\begin{align}
	&E^\rr{tot} (\bu) = E^\rr{intra} (\bu) + E^\rr{inter} (\bu) = \sum_{i=1}^3 \int_{\Omega_i} \Phi_{i, \omega} (\bu)\,d\omega\nonumber\\
&\quad=\sum_{i=1}^3 \int_{\Omega_i} \intra\,d\omega+\sum_{i=1}^3 \int_{\Omega_i} \inter\,d\omega.\label{eq:sample}
\end{align}
We then minimize $E^\rr{tot}(\bu)$ to obtain the relaxation vector in configuration space $\bu$. We will provide the form of $E^\rr{intra}$ and $E^\rr{inter}$ in Sections.~\ref{sec:intra} and \ref{sec:inter} respectively.

\subsection{Intralayer Energy}\label{sec:intra}
We proceed to describe the intralayer energy contribution in configuration space.
From linear elasticity theory, the intralayer energy due to lattice relaxation depends on the real space gradient of displacement vectors,
\begin{align}
E^\rr{intra}(\bu) &=
\sum_{i=1}^3\int_{\Omega_i}\intra  d\,\omega,
\end{align}
where $\omega=(\bm{b}^{(j)},\bm{b}^{(k)})\in \Omega_i$, and
\begin{align}
\intra = \frac{1}{2} \gradx  \widehat \bu^{(i)} (\omega):C:\gradx \widehat \bu^{(i)} (\omega).
\label{eqn:elasticity}
\end{align}
In Eq.~(\ref{eqn:elasticity}),  $C$ is the linear elasticity tensor.
For in-plane deformation, $C$ is a rank 4 tensor with its components defined as follows:
\begin{align}
	&C_{11ij} = \begin{pmatrix}
		K+G & 0 \\
		0 & G
	\end{pmatrix} \quad
	C_{12ij} = \begin{pmatrix}
		0 & K-G \\
		G & 0
	\end{pmatrix} \quad
	 \\
	&C_{21ij} = \begin{pmatrix}
		0 & G \\
		K-G & 0
	\end{pmatrix} \quad
	C_{22ij} = \begin{pmatrix}
		G & 0 \\
		0 & K+G
	\end{pmatrix},\nonumber
\end{align}
where $G$ and $K$ are the shear and bulk modulus of a monolayer (anisotropic models can be used for anisotropic 2D materials such as black phosphorous). These values can be obtained from first-principles total-energy calculations based on density functional theory (DFT) by isotropically straining and compressing the monolayer and performing a quadratic fit of the total energy as a function of the applied strain or shear. In order to incorporate out-of-plane relaxation, the elastic tensor would be a $3 \times 3 \times 3 \times 3$ tensor~\cite{dai2016twisted}, which is beyond the scope of this work.
We can compute the real space gradient of the interpolated displacement $\widehat{\bu}^{(i)}$  at an arbitrary configuration $\omega\in \Omega_i$ by the chain rule applied to Eqs.~(\ref{eqn:bj}),(\ref{eqn:bji}), and (\ref{eq:hatu}):
\begin{equation}
	 \gradx \widehat \bu^{(i)} (\omega)
= \sum_{i \neq j} \grad_{\bm{b}^{(j)}} \bu^{(i)} \cdot (A_j A_i^{-1} - I).
\end{equation}
Defining the following matrix,
\begin{equation}
	M_{ji} = A_j A_i^{-1} - I,
\end{equation}
we then have the Fourier representation of $\widehat \bu^{(i)}:$
\begin{align}
\gradx &\widehat \bu^{(i)}(\bm{b}^{(j)}, \bm{b}^{(k)}) =i \sum_{\bm{G}^{(j)}}\sum_{\bm{G}^{(k)}} \e^{i \bm{G}^{(j)} \cdot \bm{b}^{(j)}} \e^{i\bm{G}^{(k)} \cdot \bm{b}^{(k)}}\nonumber \\
&\times \tilde{\bu}^{(i)} (\bm{G}^{(j)}, \bm{G}^{(k)})
 \otimes \left[  M_{ji}^T\bm{G}^{(j)}+  M_{ki}^T\bm{G}^{(k)} \right],
\end{align}
where $\bm{G}^{(i)}$ are reciprocal lattice vectors for $L_i$ and the Fourier coefficients $\tilde{\bu}^{(i)}$ are defined according to
\begin{align}
	\bu^{(i)} (\bm{b}^{(j)},\bm{b}^{(k)}) &=\sum_{\bm{G}^{(j)},\,\bm{G}^{(k)}}  \tilde{\bu}^{(i)} (\bm{G}^{(j)},\,\bm{G}^{(k)}) \nonumber \\
	 & \quad \times \e^{i (\bm{G}^{(j)} \cdot \bm{b}^{(j)}+\bm{G}^{(k)} \cdot \bm{b}^{(k)})},
\end{align}
for $j,k\ne i$ and $j<k$.
The contribution of the mode $(\bm{G}^{(j)}, \bm{G}^{(k)})$ to the variational elastic energy is
\begin{align}
	&\mathcal{E}_i (\bm{G}^{(j)}, \bm{G}^{(k)}) \nonumber\\& =  \frac{1}{2}\left[   \tilde{\bm{u}}^{(i)}(\bm{G}^{(j)}, \bm{G}^{(k)})  \otimes \left( M_{ji}^T \bm{G}^{(j)} +M_{jk}^T \bm{G}^{(k)} \right) \right]  : C : \nonumber \\
	 &\qquad \left[    \tilde{\bm{u}}^{(i)} (\bm{G}^{(j)}, \bm{G}^{(k)}) \otimes  \left( M_{ji}^T \bm{G}^{(j)} +M_{jk}^T \bm{G}^{(k)} \right) \right]. \label{eq:fourier}
\end{align}
In terms of the $\mathcal{E}_i$'s, the total intralayer energy of the trilayer system is
\begin{align}\label{eq:intra}
	E^\rr{intra}(\bu)
= \sum_{i=1}^3 \sum_{\bm{G}^{(j)}, \bm{G}^{(k)} }    \mathcal{E}_i (\bm{G}^{(j)}, \bm{G}^{(k)}).
\end{align}

Each layer in the trilayer system has a four-dimensional configuration space $\Omega_i,$ but the in-plane interactions only span a two-dimensional submanifold. The Euler-Lagrange PDE for the
intralayer strain energy is thus not properly elliptic, but it is nonetheless nonsingular since
$\mathcal{E}_1 (\bm{G}^{(2)}, \bm{G}^{(3)})=0$ if and only if $\tilde{\bm{u}}_1 (\bm{G}^{(2)}, \bm{G}^{(3)} )=0$ or $\bm{G}^{(2)}=\bm{G}^{(3)}=0$, and similarly for $\mathcal{E}_2 (\bm{G}^{(1)}, \bm{G}^{(3)})$ and $\mathcal{E}_3 (\bm{G}^{(1)}, \bm{G}^{(2)})$.
To see this, we take $i = 1$ in Eq.~(\ref{eq:intra}) and $\mathcal{E}_1 (\bm{G}^{(2)}, \bm{G}^{(3)})  \propto  |M_{21}^T \bm{G}^{(2)} + M_{31}^T \bm{G}^{(3)}|^2$. We note that
\begin{equation}
 M_{21}^T \bm{G}^{(2)} + M_{31}^T \bm{G}^{(3)}= {\bm{G}}'^{(1)}-\bm{G}^{(2)}-\bm{G}^{(3)},
\end{equation}
where ${\bm{G}'}^{(1)}=A_1^{-T}A_2^T\bm{G}^{(2)}+A_1^{-T}A_3^T\bm{G}^{(3)}$ is in the reciprocal lattice of $L_1$, and $ {\bm{G}'}^{(1)}-\bm{G}^{(2)}-\bm{G}^{(3)}=0$ if and only if $ {\bm{G}'}^{(1)}=\bm{G}^{(2)}=\bm{G}^{(3)}=0$ by the incommensurability of the trilayer \cite{cazeaux2018energy}.  Observe that $|{\bm{G}'}^{(1)}-\bm{G}^{(2)}-\bm{G}^{(3)}|^2$ and hence $\mathcal{E}_1 (\bm{G}^{(2)}, \bm{G}^{(3)})$ can be small even though $\bm{G}^{(2)}$ and $\bm{G}^{(3)}$ are large, which is contrary to the ellipticity condition that $ \mathcal{E}_1 (\bm{G}^{(2)}, \bm{G}^{(3)})$ is a uniformly positive definite quadratic form in $(\bm{G}^{(2)}, \bm{G}^{(3)} )$ for all Fourier coefficients of the relaxation vectors $\tilde{\bm{u}}^{(1)} (\bm{G}^{(2)}, \bm{G}^{(3)} ).$
Physically, nonellipticity implies the instability due to the resonance of long wavelength modes in an infinite system. However, note that a finite physical system does not support such an instability if the elastic energy Eq.~\eqref{eq:fourier} is simply positive. In contrast, the PDEs for the intralayer energy of bilayer vdW heterostructures are elliptic. For bilayers, the variational elastic energy of $L_i$ takes the following form:
 \begin{align}
 \mathcal{E}_i (\bm{G}^{(j)}) & = \frac{1}{2}   \left(   \tilde{\bm{u}}^{(i)} (\bm{G}^{(j)}) \otimes  M_{ji}^T \bm{G}^{(j)}  \right) : C : \\ \nonumber &\qquad \left(   \tilde{\bm{u}}^{(i)} (\bm{G}^{(j)}) \otimes  M_{ji}^T \bm{G}^{(j)}  \right).
 \end{align}
 Similarly, taking $i = 1$, we have $\mathcal{E}_1 (\bm{G}^{(2)})  \propto |M_{21}^T \bm{G}^{(2)}|^2 = |\bm{G}'^{(1)} - \bm{G}^{(2)}|^2 $, where $\bm{G}'^{(1)}  = A_1^{-T} A_2^T \bm{G}^{(2)}$. In this case, $|\bm{G}'^{(1)} - \bm{G}^{(2)}|$ is bounded from zero and takes its minimum values when $\bm{G}'^{(1)} - \bm{G}^{(2)}$ are one of the columns of
$2\pi (A_1^{-T} - A_2^{-T})$, which can be understood as the primitive vectors of the bilayer moir\'e supercell reciprocal lattice.

\subsection{Interlayer Energy}\label{sec:inter}

\begin{table*}
\begin{ruledtabular}
\caption{Equilibrium lattice constant of the primitive unit cell $a_0$ (in \AA), unit cell area $\cal{A}$ (in \AA$^2$), elastic constants $K$ and $G$, and GSFE Fourier components for graphene, parallel and anti-parallel $\mathrm{WSe_2}$. Units of elastic constants and GSFE components are in meV per unit cell. }
\label{tab:values}
\begin{tabular}{@{}c|cccccccccc@{}}
System                     & $a_0\,(\mathrm{\AA}$) & $\cal{A}\, (\mathrm{\AA^2}$) & $K$   & $G$   & $c_0$  & $c_1$  & $c_2$  & $c_3$  & $c_4$ & $c_5$ \\ \midrule \cline{1-11}
Graphene                   & 2.47                  &     5.28      & 69518 & 47352 & 6.832  & 4.064  & -0.374 & -0.095 & 0.000 & 0.000 \\
Parallel $ \mathrm{ WSe_2}$  & 3.28 & 9.32 & 47123 & 31153 & 27.85& 14.08 & -2.442 & -0.773 & 0.000 & 0.000  \\
Anti-parallel $\mathrm{WSe_2}$   & 3.28 & 9.32 & 47123& 31153 & 35.03 & 12.17 & -1.899 & -0.1501 & 3.865 & 0.542 \\
\end{tabular}
\end{ruledtabular}
\end{table*}

\begin{figure}[ht!]
\centering
		\includegraphics[width=\linewidth, valign=c]{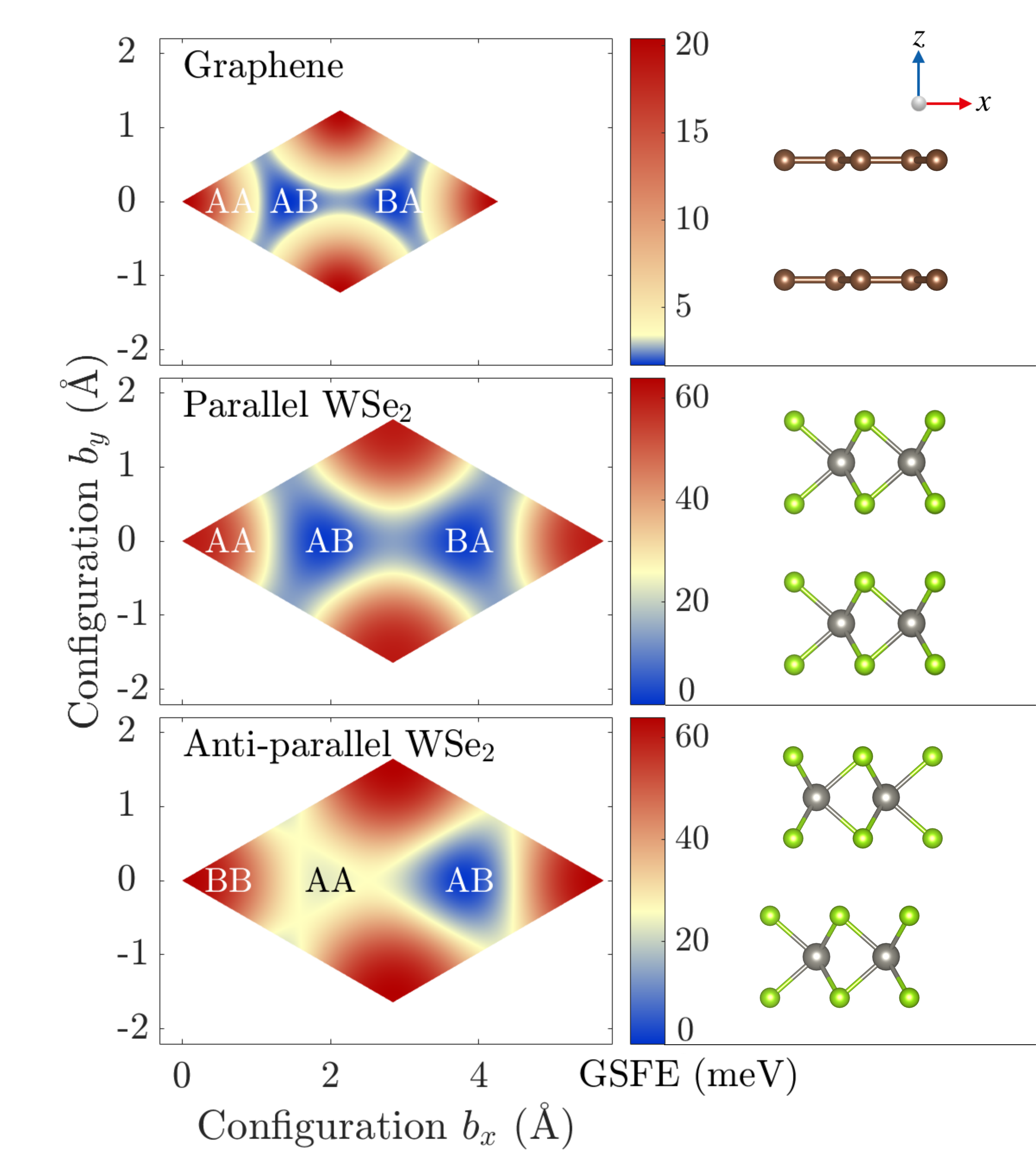}
	\caption{Left: the GSFE for bilayer graphene, parallel $\mathrm{WSe_2}$, and anti-parallel $\mathrm{WSe_2}$. Right: side views of the AA stacking orientation for each bilayer corresponding to the three cases on the left.}
	\label{fig:gsfe}
\end{figure}

We now describe the interlayer energy due to the misfit from the twist of the relaxation displacement. The origin of the interlayer energy is the slight misalignment between the two layers. This interaction is represented by the Generalized Stacking Fault Energy (GSFE). The GSFE is obtained with DFT by calculating the difference in total energy between configurations that involve relative rigid shifts between the two layers of a bilayer, with the zero corresponding to the reference ground-state-energy configuration. The GSFE can be expressed as a Fourier sum of the first few terms~\cite{kaxiras1993free}, and its functional form is given in Appendix~\ref{sec:gsfe}. The computational details of the GSFE are provided in Appendix~\ref{sec:dft} and the Fourier coefficients for graphene and $\mathrm{WSe_2}$ are provided in Table~\ref{tab:values}. The GSFE for these two materials are shown in Fig.~\ref{fig:gsfe}. We note that due to a lower symmetry, for $\rm{WSe_2}$, there are two possible stackings of a bilayer, referred to as parallel and anti-parallel, which denote the relative orientation of the two layers in the reference configuration. In graphene and parallel $\rm{WSe_2}$, the origin (AA stacking) is when the two layers are exactly aligned, and AB/BA stacking is when one sublattice is aligned with the other. In anti-parallel $\rm{WSe_2}$, we refer to the origin (BB stacking) as the stacking configuration when the chalcogen atoms (Se) are aligned, the AA stacking order as when the metal atoms (W) are aligned, and the AB stacking order is when the metal atom is aligned with the chalcogen atom. The stacking configuration is consistent with the notation in~\citet{philips2019commensurate}.

Unrelaxed localized misfit site energies $\Phi^{\rr{misfit}}_{i \pm}$: $\Gamma^{i\pm 1} \mapsto \mathbb{R}$ are defined by
\begin{equation}
	\misfit{i+}{i}{i+1} = \misfit{(i+1)-}{i+1}{i}.
\end{equation}
These site energies thus satisfy the following symmetry relation for $\bm{s} \ \in\  A_i^{-1} \Gamma_i = A_{i+1}^{-1} \Gamma_{i+1} \equiv [0, 1)^2:$
\begin{equation}\label{eq:GSFESym}
    \Phi^\mathrm{misfit}_{(i+1)-}\left (A_i \bm{s} \right ) = \Phi^\mathrm{misfit}_{i+} \left (-A_{i+1} \bm{s} \right ),
\end{equation}
and we see that the misfit energies $\Phi^\mathrm{misfit}_{(i+1)-}$ and $\Phi^\mathrm{misfit}_{i+}$ can be computed from the same misfit energy between $L_{i}$ and $L_{i+1}$. As mentioned above, the simulations in this paper approximate the misfit energy $\Phi^{\mathrm{misfit}}_{i\pm}$ by the GSFE, $V^{\mathrm{GSFE}}_{i\pm}$, defined in Eq.~(\ref{eqn:vgsfe}) in Appendix~\ref{sec:gsfe}.

When the layers are relaxed, the argument of $V^\rr{GSFE}$ is modified to be the relaxation-modulated local shift $\bm{B}^{i \rightarrow j}$ defined in Eq.~(\ref{eqn:local_shift}). The interlayer contribution of each layer can be given by
\begin{equation}
\inter =
	\begin{cases}
		\frac{1}{2} \misfitr{1+}{1}{2},  \hfill \rr{for}\ \omega\in \Omega_1, \\
		\frac{1}{2} \left[ \misfitr{2+}{2}{3} + \misfitr{2-}{2}{1} \right], \hfill \\ \hfill \rr{for}\ \omega\in \Omega_2, \\
		\frac{1}{2} \misfitr{3-}{3}{2}, \hfill \rr{for}\ \omega\in \Omega_3.
	\end{cases}
\end{equation}
The total interlayer energy can therefore be obtained by integrating over all local configurations $\omega$:
\begin{align}
	&E^\mathrm{inter} = \frac{1}{2}\int_{\Omega_1}\misfitr{1+}{1}{2}\,\mathrm{d}\omega  \\
&\quad+\frac{1}{2}\int_{\Omega_2}\left[ \misfitr{2-}{2}{1} + \misfitr{2+}{2}{3} \right]\,\mathrm{d}\omega \nonumber \\
&\quad+\frac{1}{2}\int_{\Omega_3} \misfitr{3-}{3}{2} \,\mathrm{d}\omega.\nonumber
\end{align}
Note that for hetero-trilayer materials, $V^{\mathrm{GSFE}}_{2-}$ and $V^{\mathrm{GSFE}}_{2+}$ have different functional forms.

\begin{figure}[ht!]
	\centering
	\begin{subfigure}{\linewidth}
	\centering
	\includegraphics[width=\textwidth, valign=c]{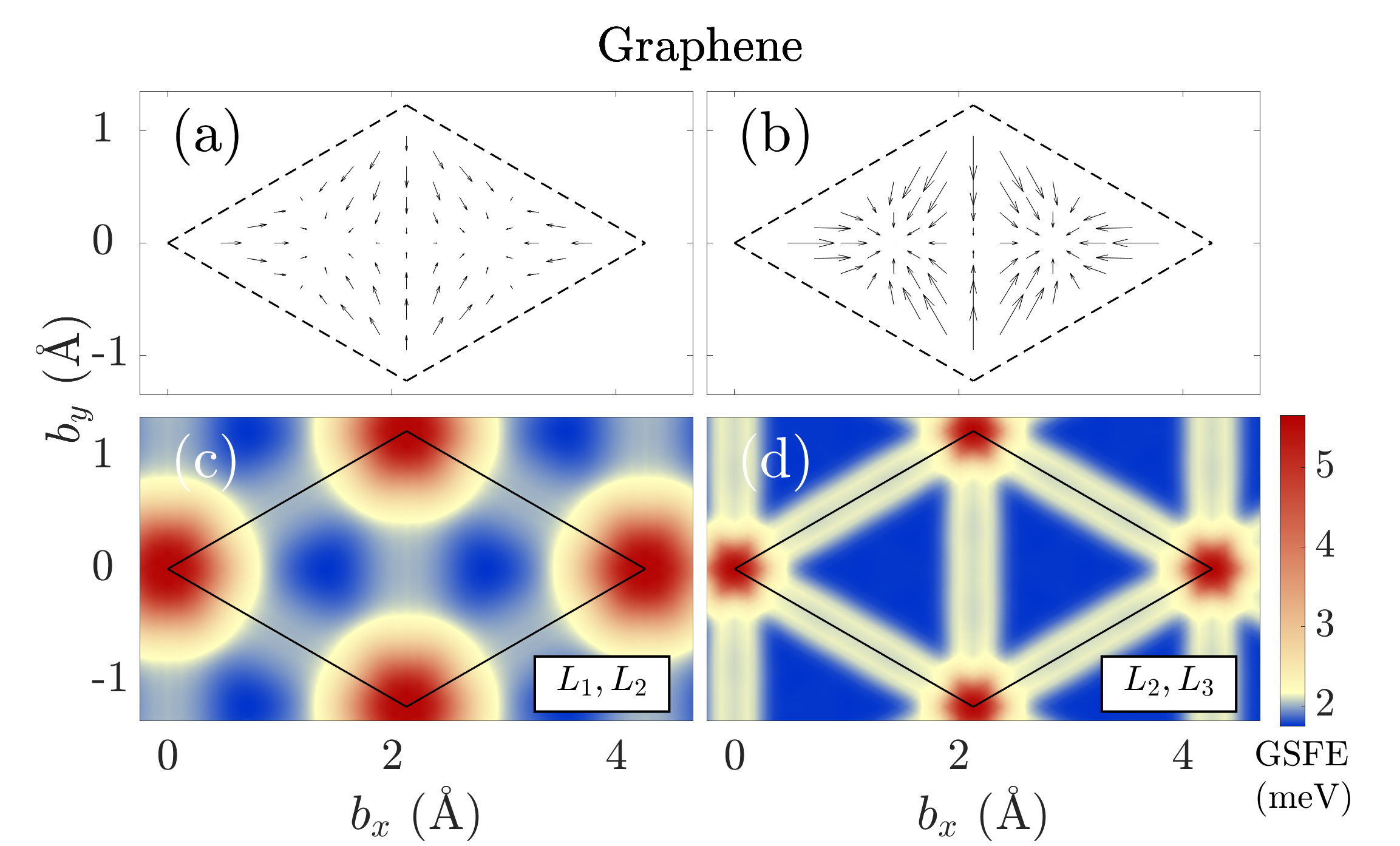}
	\end{subfigure}
	\begin{subfigure}{\linewidth}
	\centering
	\includegraphics[width=\textwidth, valign=c]{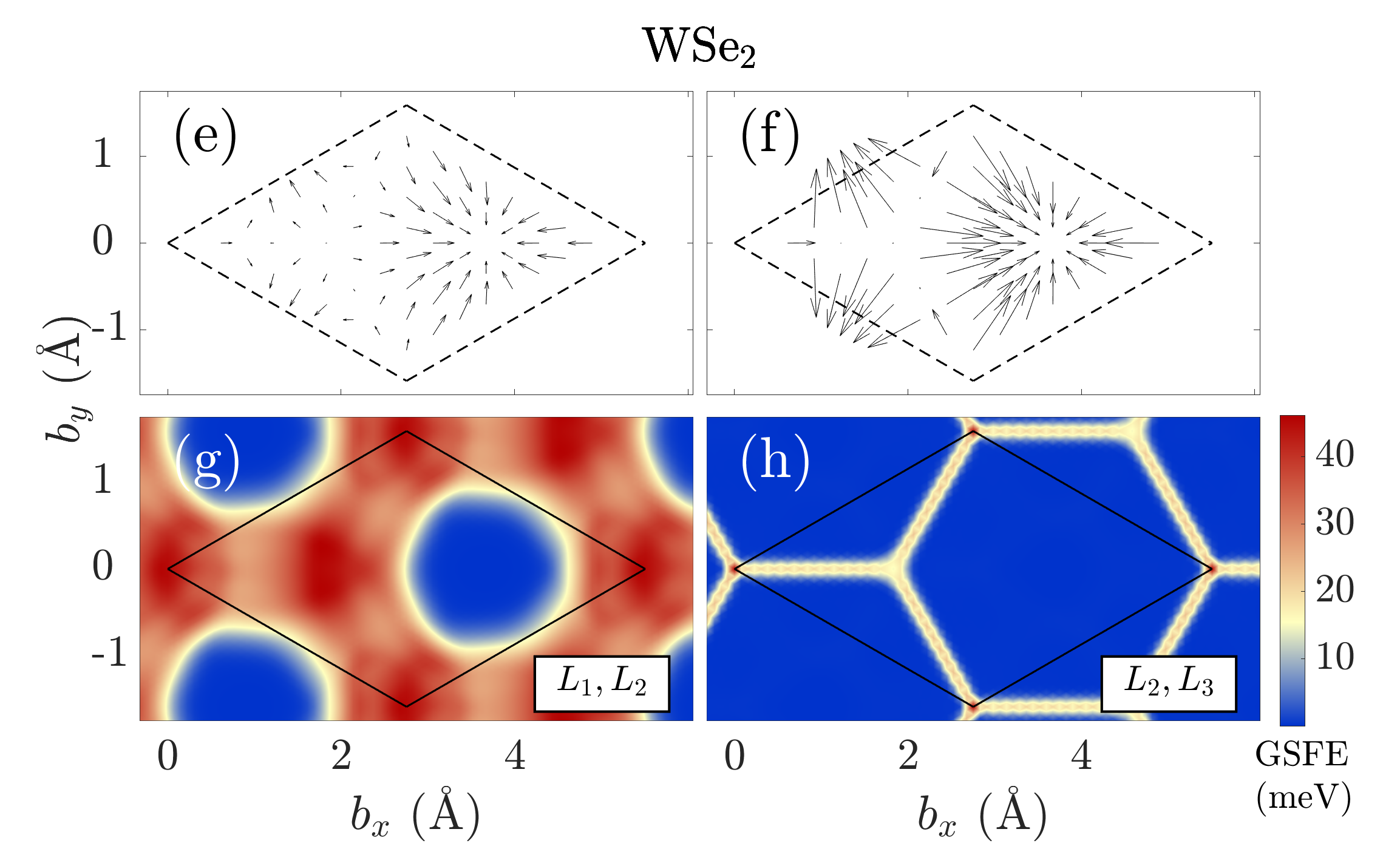}
	\end{subfigure}%
\caption{Two-dimensional line cuts of the relaxation pattern and relaxed GSFE in configuration space for (a)-(d) trilayer graphene and (e)-(h) trilayer anti-parallel $\mathrm{WSe_2}$ at $\theta_{12} = 0.5^\circ, \theta_{23} = 0.2^\circ$. (a) and (b) show the change in the local shift vectors, $(\bm{B}^{i\rightarrow2} - \bm{b}^{i\rightarrow2})$. (a) shows $(\bm{B}^{1\rightarrow2} - \bm{b}^{1\rightarrow2})$, averaged over the $L_3$ configuration, and (b) shows $(\bm{B}^{3\rightarrow2} - \bm{b}^{3\rightarrow2})$, averaged over the $L_1$ configuration. (c) and (d) show the relaxed GSFE. (c) shows the sum of the GSFE between the $L_1, L_2$ bilayer pair, averaged over the $L_3$ configuration, and (d) shows the sum of the GSFE between the $L_2, L_3$ bilayer pair, averaged over the $L_1$ configuration. (e)-(h) are the corresponding plots for anti-parallel ${\mathrm WSe_2}$.}
\label{fig:config_results}
\end{figure}

\subsection{Relation between Real Space and Configuration Space}\label{sec:relation}
Finally, we describe the mapping between real space and configuration space.
Each adjacent pair of layers $i$ and $j$ forms a moir\'e supercell, $\Gamma^{ij},$ given by the columns of the matrix 
$(I - A_i A_j^{-1})^{-1} A_i=\left(A_i^{-1}-A_j^{-1}\right)^{-1}$. At the real space position at a lattice site in $L_i$,
$\bm{R}^{(i)}_{mn}$,
we can calculate the local configuration relative to $L_j$ by calculating the local shift vector (see Eq.~(\ref{eqn:bij_site})):
\begin{align}\label{eq:config}
\bm{b}^{(j)} (\bm {R}_{mn}^{(i)})& =
\bm{R}_{mn}^{(j)} - \bm {R}_{mn}^{(i)}
=(A_j - A_i)
\begin{pmatrix}
m \\ n
\end{pmatrix}  \nonumber \\&=  (A_jA_i^{-1}-I)\bm {R}_{mn}^{(i)}\in\Gamma^{(j)}.
\end{align}
Note that this construction agrees with evaluating the interpolated disregistry $\bm{b}^{(j)}$ given by Eq.~(\ref{eqn:bj}) at a lattice site $\bm{R}^{(i)}_{mn}$ as the reference point.
This result can be generalized to any real space position $\bm r$:
\begin{align}
\bm{b}^{i\rightarrow j}(\omega(\bm r))=(A_jA_i^{-1}-I)\bm r  \in\Gamma^{(j)},
\end{align}
where we notice that $\bm{b}^{i \rightarrow j} (\bm r)$ is periodic in the bilayer moir\'e supercell $\Gamma^{ij}$ in the continuum approximation, meaning it sends a position in the bilayer supercell to the unit cell of $L_j$. This generalization is equivalent to evaluating the interpolated disregistry given in Eq.~(\ref{eqn:bji}) at a configuration $\omega$.

The relation between the displacement
defined in real space, $\bm{U}^{(i)}(\bm r)$, and in configuration space, $\bm{u}^{(i)}(\bm r)$, can then be given by evaluating $\widehat{\bm{u}}^{(i)}(\omega)$ at $\bm{b}^{i\to j} (\bm r)$ and $\bm{b}^{i\to k} (\bm r)$ with Eq.~(\ref{eq:config}) above at a position $\bm r$
to obtain
\begin{equation}\label{eq:displacement}
\bm{U}^{(i)}(\bm r)= \bm{u}^{(i)}(\bm{b}^{i\rightarrow j}(\omega(\bm r)),\bm{b}^{i\rightarrow k}(\omega(\bm r))),
\end{equation}
where $j,k\ne i$ and $j<k.$
We see that $\bm{U}^{(i)}(\bm r)$ is not generally periodic since $\bm{b}^{(j)}(\bm r)$ and $\bm{b}^{(k)}(\bm r)$ are each
periodic on the incommensurate moir\'e supercells $\Gamma^{ij}$ and $\Gamma^{ik}.$
In comparison, for incommensurate bilayers $\bm{U}^{(i)}(\bm r)=\widehat{\bm{u}}^{(i)}(\omega)$ is periodic on $\Gamma^{ij}.$

\section{Results}\label{sec:results}
\subsection{General Twist Angles, $\theta_{12} \neq \theta_{23}$}
\begin{figure*}[ht!]
	\centering
	\includegraphics[width=\linewidth, valign=c]{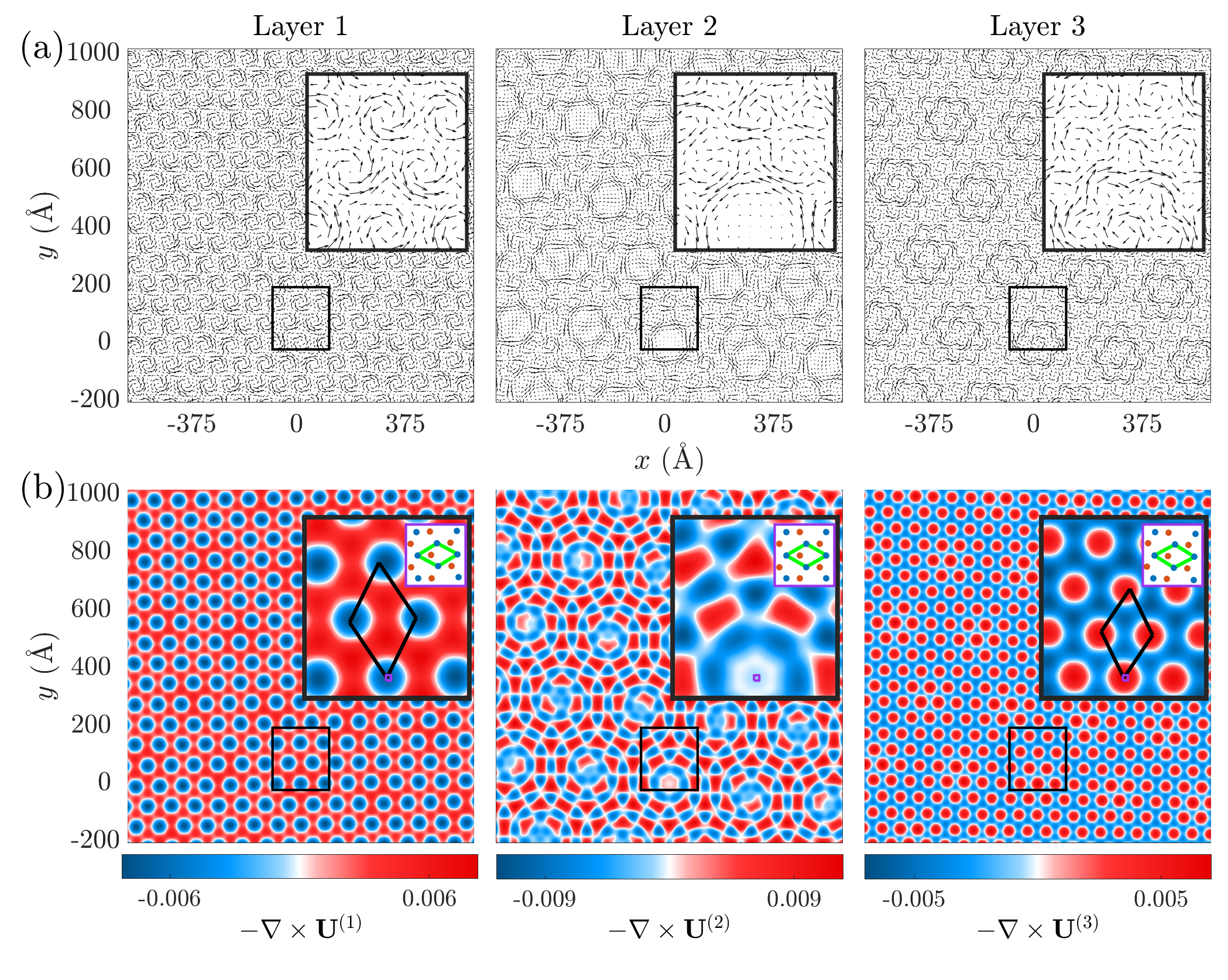}
	\centering
\caption{Real space relaxation pattern for twisted trilayer graphene with $\theta_{12} = 1.73^\circ$, $\theta_{23} = 2.24^\circ$, using discretization $N = 81$. Insets on the upper right corner zoom in on the black boxed areas near the origin. (a) the relaxation displacement $\bm{U}^{(i)} (\bm r)$.  (b) Color plot shows the curl of the displacement vectors $-\grad_{\bm r} \times \bm{U}^{(i)}(\bm r)$ indicating the rotation direction: blue is clockwise rotation and red is counterclockwise rotation, and white means no rotation. Black parallelograms on $L_1$ and $L_3$ show the approximate supercell vectors that correspond to  $\theta_{12}$ and $\theta_{23}$ respectively, calculated using Eq.~\eqref{eqn:asc}. The small insets on the upper right corner zoom in on the small purple boxed areas in the black inserts. In the purple insets, blue and red dots show A and B sublattices respectively, and the green parallelogram shows the monolayer cell.   }
\label{fig:realspace}
\end{figure*}

\begin{figure*}[ht!]
	\centering
	\includegraphics[width=\linewidth, valign=c]{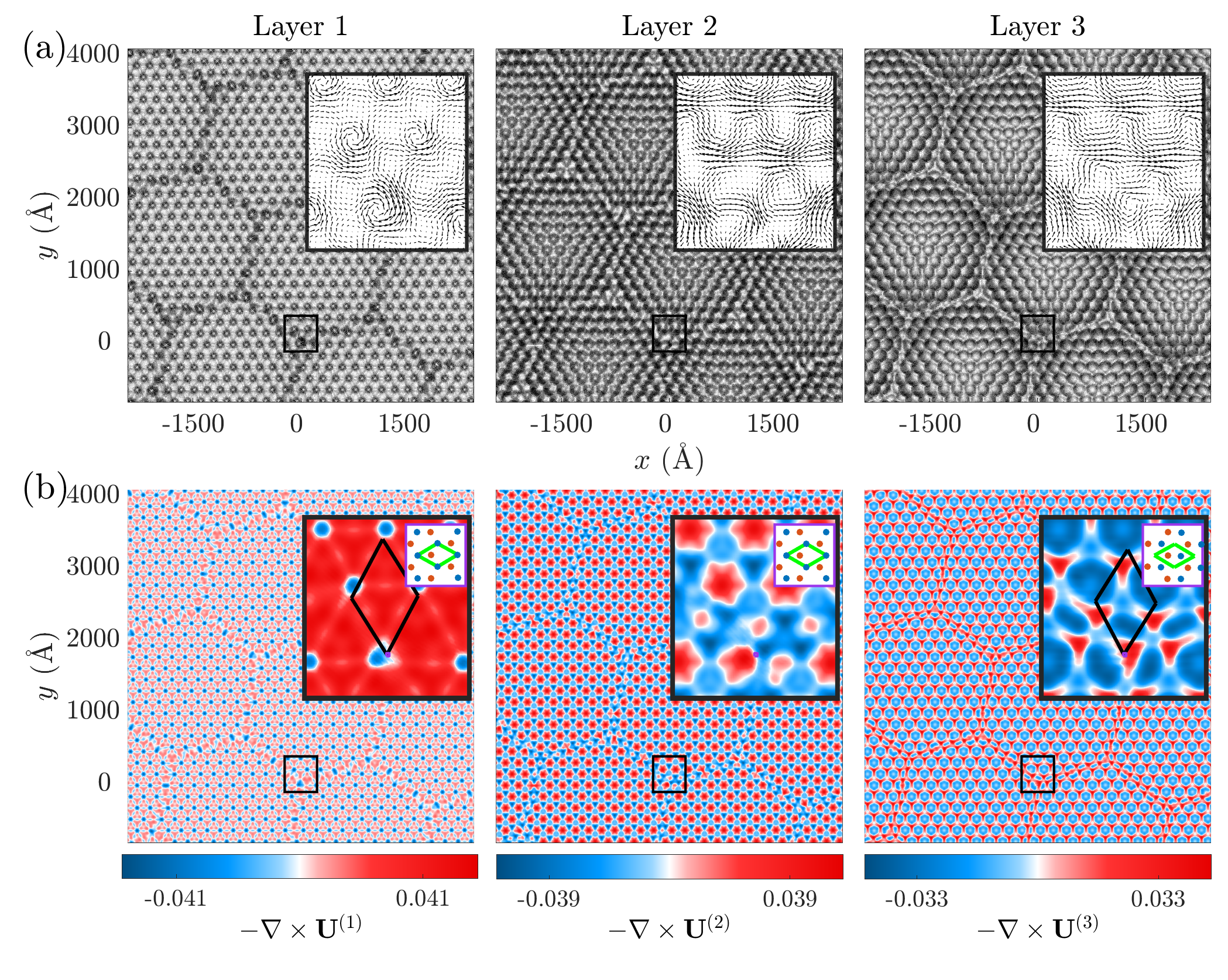}
	\centering
\caption{Same as Fig.~\ref{fig:realspace}, for trilayer $\mathrm{WSe_2}$, with parallel-stack between $L_1$ and $L_2$ at $\theta_{12} = 1.0^\circ$, and anti-parallel-stack between $L_2$ and $L_3$ at $\theta_{23} = 1.1^\circ$.}
\label{fig:realspace_wse2}
\end{figure*}

We perform energy minimization in configuration space with the discretized grid $N \times N \times N \times N$ using the \texttt{JULIA OPTIM} package ~\cite{mogensen2018optim}.  We present our results in Fig.~\ref{fig:config_results} for $\theta_{12} = 0.5^\circ, \theta_{23} = 0.2^\circ$  by taking two-dimensional line cuts of the configuration space, using $N = 36$. We show the results for twisted trilayer graphene in Fig.~\ref{fig:config_results}a-(d) and for trilayer anti-parallel $\mathrm{WSe_2}$ in Fig.~\ref{fig:config_results}(e)-(h). For the $L_1, L_2$ bilayer pair, we take the average over the configurations of $L_3$ (left), and similarly for the $L_2, L_3$ bilayer pair, we take the average over the configurations of $L_1$ (right). For the relaxation displacement vectors, we show the change in modulated local shift vector compared to the local shift vector without relaxation between the bilayer pair, i.e., $(\bm{B}^{i\rightarrow2} - \bm{b}^{i\rightarrow2}) $ for $i = 1, 3$, where $\bm{b}^{i\rightarrow2}$ and $\bm{B}^{i\rightarrow2}$ are the local shift vector and the modulated local shift vector respectively, given in Eqs.~(\ref{eqn:bji}) and (\ref{eqn:local_shift}) respectively. For example, for the relative relaxation between the $L_1$ and $L_2$ pair, we plot the following quantity:
\begin{align}
&\bm{B}^{1\rightarrow2}(\bm{b}^{(2)},\bm{b}^{(3)}) - \bm{b}^{1\rightarrow2}(\bm{b}^{(2)},\bm{b}^{(3)}) \\&= \widehat{\bm{u}}^{(2)}(A_1A_2^{-1} \bm{b}^{(2)}, \bm{b}^{(3)} - A_3 A_2^{-1} \bm{b}^{(2)})  - \bm{u}^{(1)}(\bm{b}^{(2)}, \bm{b}^{(3)}),\nonumber
\end{align}
as a function of $\bm{b}^{(2)}$, averaged over the $\bm{b}^{(3)}$. Physically, this means displaying the relative relaxation vectors from $L_2$ to $L_1$, averaged over all the $L_3$ shifts. Similarly, for $L_2$ and $L_3$, we show $(\bm{B}^{3\rightarrow2} - \bm{b}^{3\rightarrow2})$ as a function of $\bm{b}^{(2)}$, averaged over all the $L_1$ configurations. The relaxation between $L_2$ and $L_3$ for both graphene and $\mathrm{WSe_2}$ is stronger because of the smaller corresponding twist angle $\theta_{23}$. The relaxation displacement vectors point towards the AB/BA stacking in graphene and AB stacking in anti-parallel $\mathrm{WSe_2}$, which are the equilibrium stacking orders.
For both graphene and $\mathrm{WSe_2}$, the relaxed GSFE exhibits domain wall formation between the adjacent bilayer pairs with different thicknesses. A smaller corresponding twist angle gives rise to stronger relaxation and thinner domain walls, similar to twisted bilayer graphene. The shape of the domains are different depending on the lattice symmetry: graphene has triangular domains while anti-parallel $\mathrm{WSe_2}$ has hexagonal domains due to its broken inversion symmetry.


\begin{figure*}[ht!]
\centering
	\begin{subfigure}{0.8\linewidth}
	\centering
	\includegraphics[width=0.8\textwidth, valign=c]{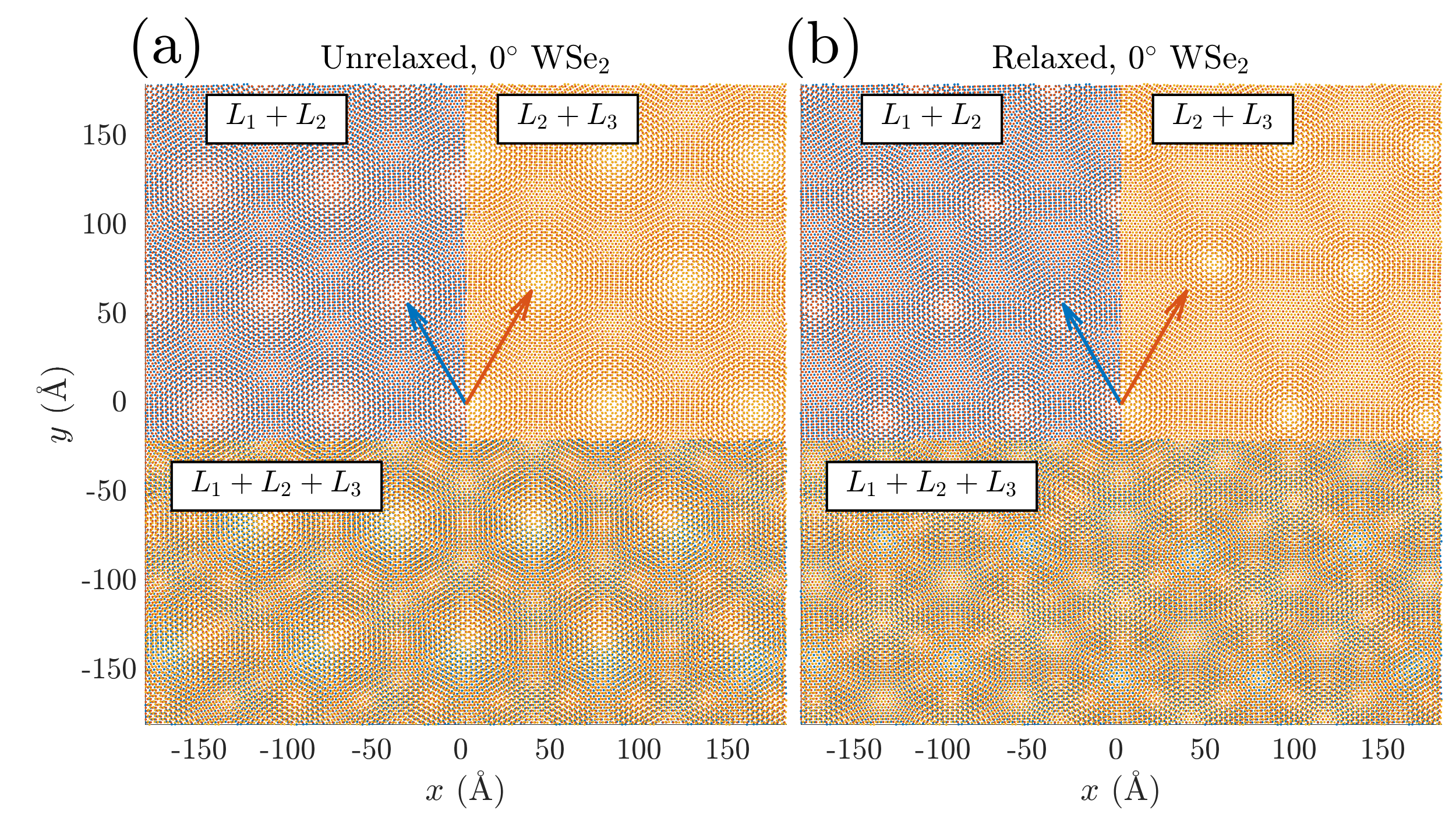}
	\end{subfigure}
	\begin{subfigure}{0.8\linewidth}
	\centering
	\includegraphics[width=0.8\textwidth, valign=c]{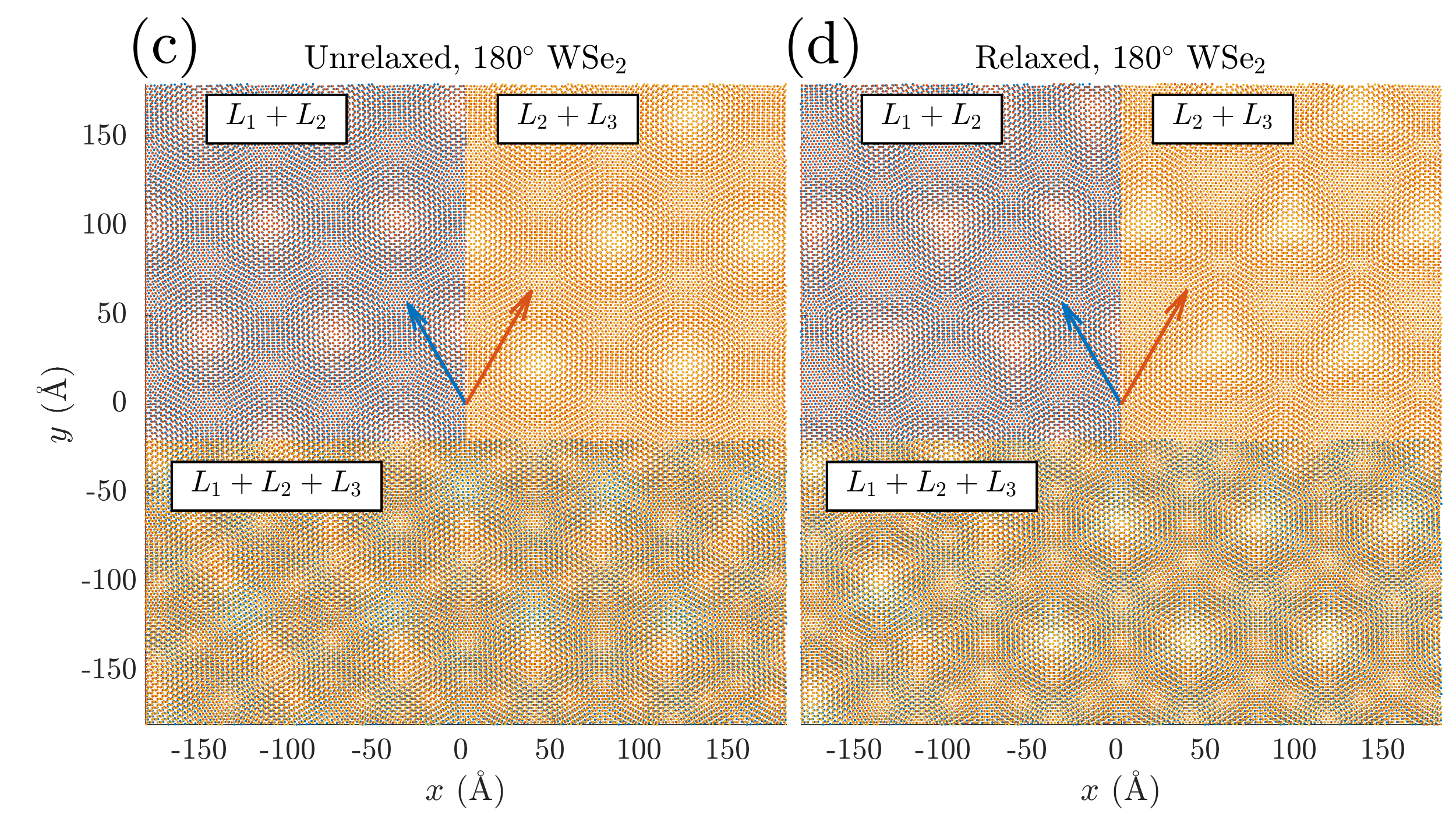}
	\end{subfigure}%
\caption{Comparison between unrelaxed and relaxed  atomic positions in twisted trilayer $\mathrm{WSe_2}$ with (a)-(b) parallel and (c)-(d) anti-parallel stackings at $\theta_{12} = 2.6^\circ$ and $\theta_{23} = 2.3^\circ$. In all panels, top left: atomic positions of $L_1$ and $L_2$ only (removing $L_3$); top right: atomic positions of $L_2$ and $L_3$ only (removing $L_1$); bottom: atomic positions of the three layers. Red and blue arrows are guides to eyes, showing the bilayer moir\'e lengths that correspond to $\theta_{12} = 2.6^\circ$ and $\theta_{23} = 2.3 ^\circ$. }
\label{fig:domain}
\end{figure*}

Unlike in the case of twisted bilayer heterostructures, there does not exist in trilayers a linear mapping at the continuous level between configuration space and a periodic moir\'{e} supercell in real space. From the energy minimization in configuration space, we obtain a set of displacement vectors, $\bu^{(i)}(\bm{b}^{(j)},\bm{b}^{(k)}),$ for each layer.
The real space displacement $\bm{U}^{(i)}(\bm{R}^{(i)})$ at lattice points $\bm{R}^{(i)}$ in $L_i$ is then given
by evaluating $\bm{U}^{(i)}(\bm{R}^{(i)})={\bm{u}}^{(i)}(\bm{b}^{(j)}(\bm{R}^{(i)}),\bm{b}^{(k)}(\bm{R}^{(i)})).$
In contrast, the configuration space for the bilayer case is two-dimensional and the displacement is more simply given by $\bm{U}^{(i)}(\bm{R}^{(i)})={\bm{u}}^{(i)}(\bm{b}^{(j)}(\bm{R}^{(i)})).$
Fig.~\ref{fig:realspace} shows the real space relaxation displacement vectors $\bm{U}^{(i)} (\bm r)$ and the curl of $-\bm{U}^{(i)} (\bm r)$, $-(\grad_{\bm r} \times \bm{U}^{(i)}(\bm r))$, which measures the rotation component of the modulation displacement vector field, for twisted trilayer graphene~\cite{zhang2018structural}. Physically, this quantity corresponds to the change of the local twist angle, which can be measured using the SQUID-on-tip technique~\cite{uri2019mapping}. In Fig. \ref{fig:realspace}a, we approximate the supercell lattice constant that corresponds to the twist angle $\theta_{i, i+1}$ according to
\begin{equation}
a_\mathrm{sc} (\theta_{i, i+1}) = \frac{a_0}{2 \sin (\theta_{i, i+1}/2)},
\label{eqn:asc}
\end{equation}
where $a_0$ is the lattice constant of the graphene/$\mathrm{WSe_2}$ unit cell given in Table~\ref{tab:values}.
The length scale $a_\mathrm{sc} (\theta_{i, i+1})$ can be understood as the periodicity of a twisted bilayer with the twist angle $\theta_{i, i+1}$ in the continuum limit.
Fig.~\ref{fig:realspace}a shows the relaxation displacement vectors. Near the origin, the relaxation tends to follow the local rotation to increase the local twist angle to reduce the size of the AA spots, which agrees with the bilayer results ~\cite{zhang2018structural}.

To better understand the relaxation pattern, Fig.~\ref{fig:realspace}b shows $-(\grad_{\bm r} \times \bm{U}^{(i)} (\bm r))$.
On the bilayer moir\'e scales (see insets on the upper right corners), the relaxation pattern has two length scales: in $L_1$, the length scale of the relaxation is dominated by the continuum supercell lattice constant of $\theta_{12}$, $a_\mathrm{sc} (\theta_{12})$; in $L_3$, the relaxation has a length scale $a_\mathrm{sc} (\theta_{23})$ instead. In the middle layer ($L_2$), the relaxation pattern is a result of the interference between the two bilayer moir\'e lengths, as a result of the coupling to both $L_1$ and $L_3$.
Zooming out, we observe structures at a larger length scale than both of the bilayer moir\'e lengths, which is the ``moir\'{e} of moir\'{e},'' due to the interference between the two bilayer moir\'e cells. Note that these larger structures are not identical to each other (most visible in Fig.~\ref{fig:realspace}b, Layer 2) due to the incommensurability of the system. The emergence of the larger length scales can also be explained by the nonellipticity of the system (see Section~\ref{sec:intra}). The reciprocal space variational elastic energy components given in Eq.~\eqref{eq:fourier}, $\mathcal{E}_i (\bm{G}^{(j)}, \bm{G}^{(k)}),$ have multipliers $|\bm{G}'^{(i)} - \bm{G}^{(j)} - \bm{G}^{(k)}|^2$ which can take an arbitrarily small value, which means the real space relaxation pattern can have arbitrarily large structures. The quasi-periodic large structures we observe in Fig.~\ref{fig:realspace} manifests the nonellipticity of the Euler-Lagrangian PDE, and since these structures are aperiodic, zooming out further will lead to the appearance of even longer quasi-periodic structures. This shows our model's capability of capturing the incommensurability of the trilayer relaxation pattern.

To show the generality of our model, we can examine the results for other materials. For example, Fig.~\ref{fig:realspace_wse2} shows the relaxation displacement vectors and the corresponding curl in trilayer $\mathrm{WSe_2}$ heterostructures, with parallel stacking between $L_1$ and $L_2$ and anti-parallel stacking between $L_2$ and $L_3$. Similarly to the graphene case, we observe the coupling between two length scales that corresponds to $\theta_{12}$ and $\theta_{23}$. On the bilayer moir\'e length scales, the relaxation forms triangular domains between $L_1$ and $L_2$ and hexagonal domains between $L_2$ and $L_3$. On the moir\'e of moir\'e length scale, the overall domains are also hexagonal, since the overall inversion symmetry of the system is broken by anti-parallel $\mathrm{WSe_2}$. The shape can also be understood as the interference pattern between the triangular domains and hexagonal domains.

We note that when $\theta_{12} = -\theta_{23}$, $L_1$ and $L_3$ are aligned with a twisted middle layer. The system can also be modeled as a bilayer system, the interlayer energy of $L_2$ doubles and the strength of relaxation in $L_2$ doubles. We used our trilayer model to calculate the relaxation pattern for such systems and obtain a similar result as the relaxation pattern shown in \citet{carr2019coexistence}.

We examine the real space atomic positions after relaxation. Fig.~\ref{fig:domain} compares the unrelaxed and relaxed atomic positions for parallel and anti-parallel $\mathrm{WSe_2}$ at $\theta_{12} = 3^\circ$, $\theta_{23} = 2.3^\circ$. Note that the unrelaxed atomic positions are different in the parallel and anti-parallel stackings (Fig.~\ref{fig:domain}a, c). In both cases, the relaxation forms domain between $L_1$ and $L_2$, as well as between $L_2$ and $L_3$, but at different length scales, to enlarge equilibrium stacking orders. Fig.~\ref{fig:domain}b shows that in parallel $\mathrm{WSe_2}$, AB/BA stacking regions are enlarged and the AA stacking regions shrink. Fig.~\ref{fig:domain}d shows the domain formation in anti-parallel $\mathrm{WSe_2}$, in which AB stacking regions are enlarged.

\subsection{Equal Twist Angles, $\theta_{12} = \theta_{23}$}
\begin{figure*}[ht!]
\centering
	\begin{subfigure}{0.7\linewidth}
	\centering
	\includegraphics[width=\textwidth, valign=c]{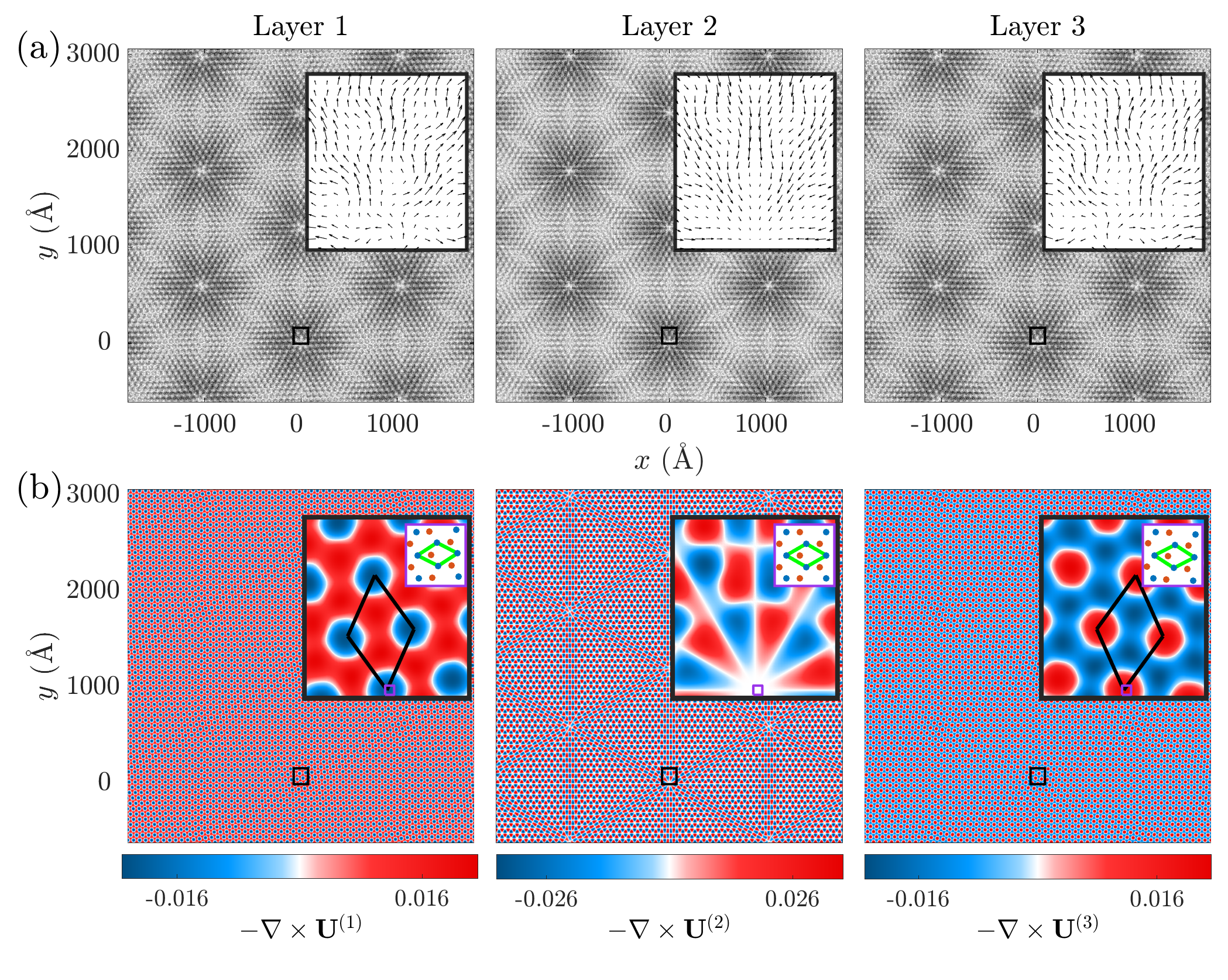}
	\end{subfigure}
	\begin{subfigure}{0.7\linewidth}
	\centering
	\includegraphics[width=\textwidth, valign=c]{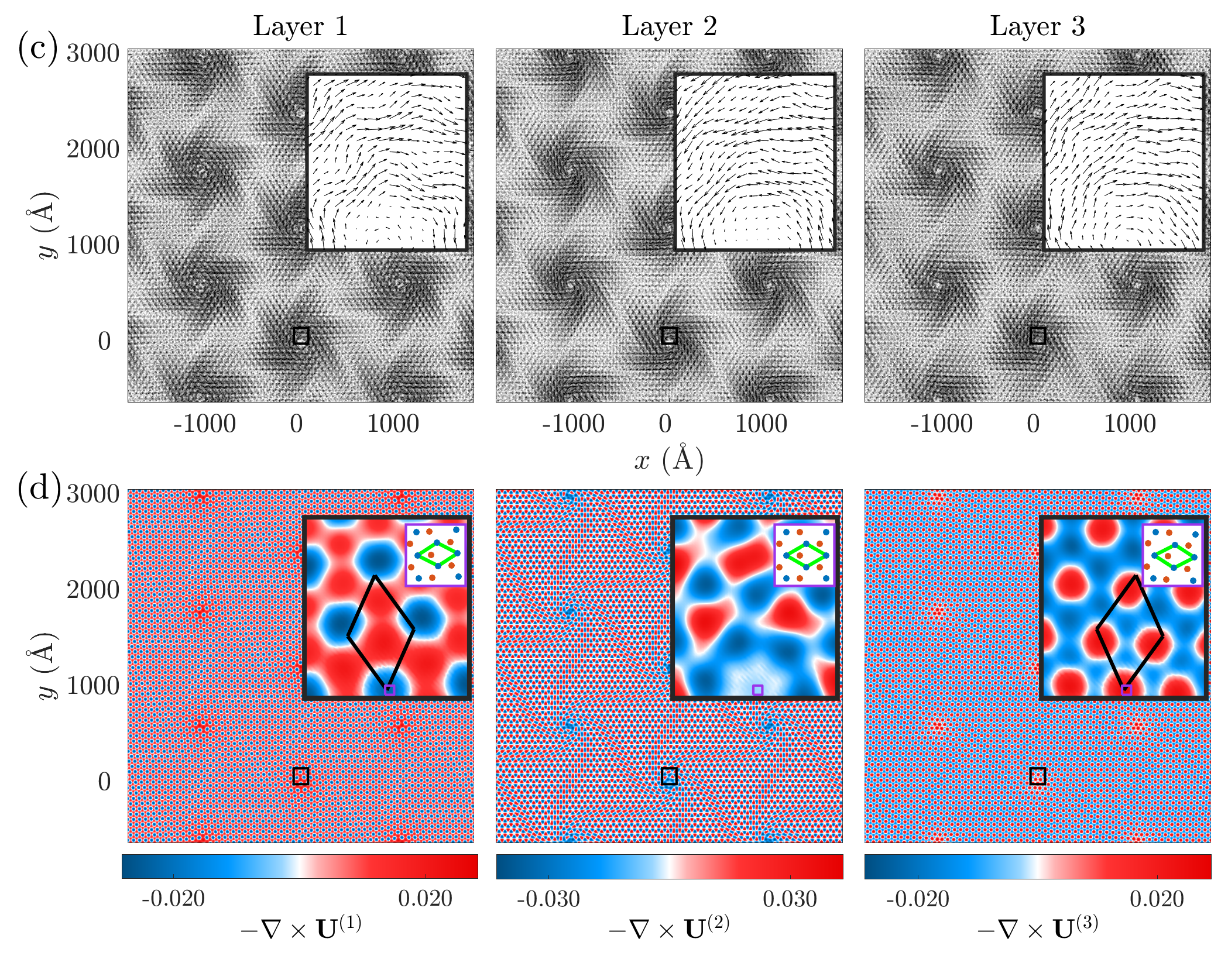}
	\end{subfigure}%
\caption{Relaxation pattern for parallel $\mathrm{WSe_2}$ at $\theta_{12} = \theta_{23} = 3^\circ$, same as Fig.~\ref{fig:realspace}, showing the symmetric solution in (a)-(b) and the asymmetric solution in (c)-(d).}
\label{fig:eq_twist}
\end{figure*}

In the special case where  $\theta_{12} = \theta_{23}$, we found that there exist two types of solutions: symmetric and asymmetric. An example is shown in Fig.~\ref{fig:eq_twist}. The symmetric solution (Fig.~\ref{fig:eq_twist}a-b) preserves the two-fold rotational symmetry along the $z$-axis (out-of-plane direction) of the unrelaxed trilayers, whereas the asymmetric solution (Fig.~\ref{fig:eq_twist}c, d) exhibits a pinwheel-shaped relaxation pattern, which remarkably breaks this symmetry. Near the origin (local AAA stacking) of $L_2$, the relaxation has zero magnitude for the symmetric solution, whereas the relaxation is finite for the asymmetric solution. The asymmetric solution is found to have a lower energy in this case: at $\theta_{12} = \theta_{23} = 3^\circ$ for parallel $\rm{WSe_2}$, the total energies per monolayer unit cell after relaxation are 39.34$\, \rm{meV}$ and 39.30$\, \rm{meV}$ for the symmetric and asymmetric solutions respectively.
Note that there exists two asymmetric solutions with equal energy, in which the pinwheel rotates in the opposite directions.

Physically, the asymmetric solution redistributes the local AA stacking density between the neighboring bilayer pairs to reduce the total AA areas and consequently the total interlayer energy, at the cost of additional elastic energy due to the pinwheel relaxation pattern. This can be seen from the insets in Fig.~\ref{fig:eq_twist}. In the insets of Fig.~\ref{fig:eq_twist}b for $L_1$ and $L_3$, the black parallelograms (bilayer moir\'e cells) agrees with the length scale of the relaxation pattern. In $L_2$, relaxation is canceled out at the AAA spots (white color near the origin) to preserve the two-fold rotational symmetry. In Fig.~\ref{fig:eq_twist}d, however, the insets show that the relaxation near the origin in $L_1$ and $L_3$ now has different length scales than the bilayer moir\'e length corresponding to $3^\circ$. We also observe that the relaxation in $L_2$ obtains a preferred net rotation, which breaks the symmetry between the top and bottom layers. These results suggest that relaxation can spontaneously break the symmetry of the system, unlike in twisted bilayers in which the in-plane relaxation always preserves the symmetry. 

A bifurcation point occurs at a critical twist angle, $\theta_c \approx 4.55^\circ$. Above $\theta_c$, the only solution is the symmetric one; below $\theta_c$, the symmetric solution becomes an unstable saddle point and the two counter-rotating asymmetric solutions become the global minima. The existence of a critical twist angle can be understood from the competition between reducing the misfit energy and increasing elastic energy. As the twist angle increases, the moir\'e of moir\'e length scale becomes smaller, and therefore elastic energy cost for the pinwheel pattern increases. Eventually, the reduction in the misfit energy is not enough to compensate for the elastic energy cost of the pinwheel pattern and the system no longer supports the asymmetric solution.

\begin{figure*}[ht!]
\centering
\centering
	\begin{subfigure}{0.4\linewidth}
	\centering
	\includegraphics[width=\textwidth, valign=c]{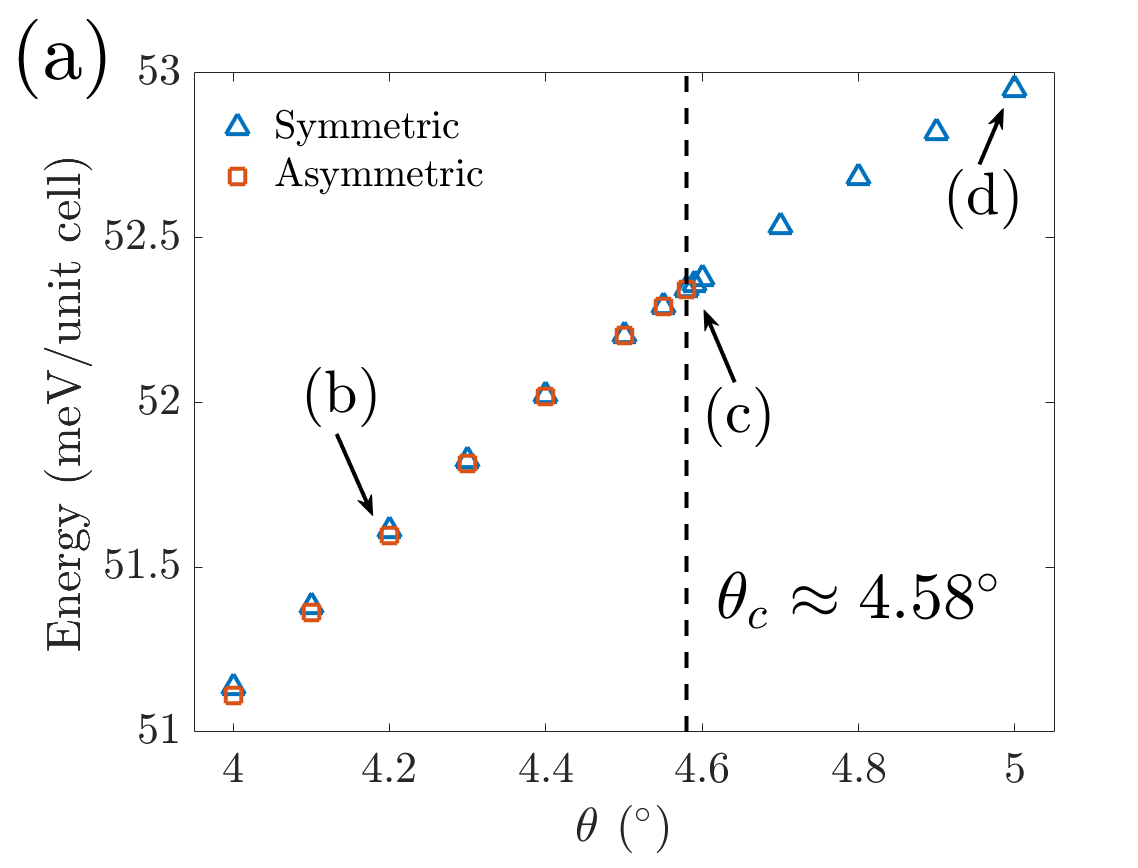}
	\end{subfigure}
	\begin{subfigure}{\linewidth}
	\centering
	\includegraphics[width=\textwidth, valign=c]{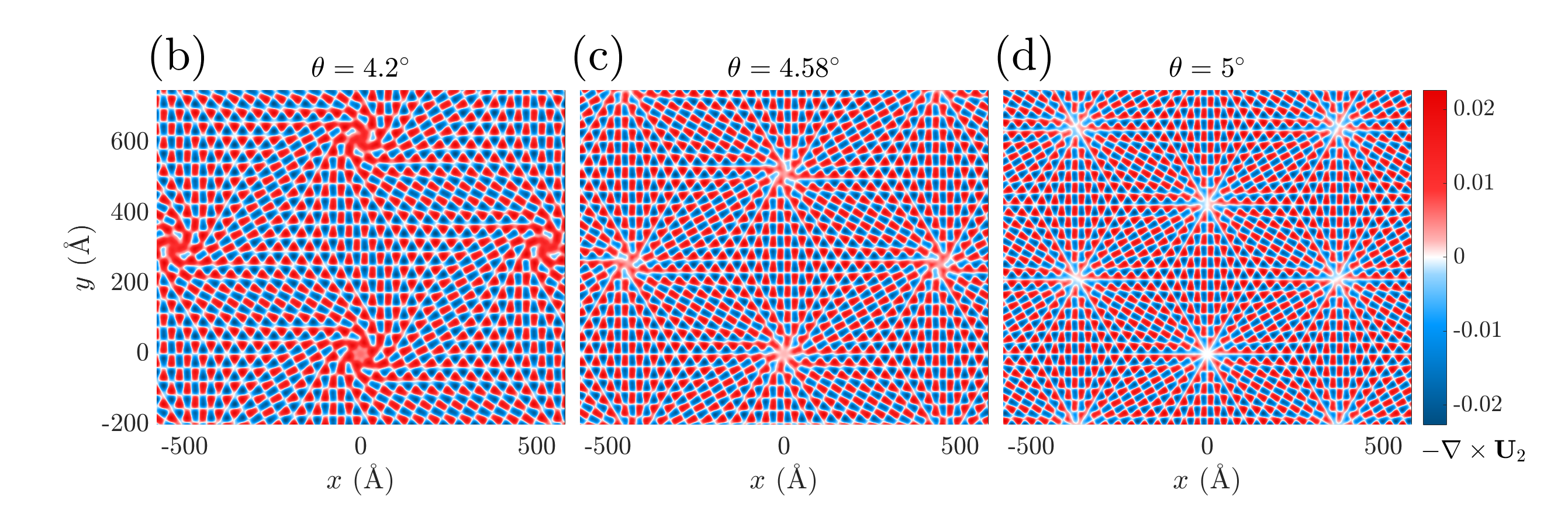}
	\end{subfigure}%
\caption{(a) Relaxed total energy as a function of the twist angle shows the bifurcation critical point is around $\theta = 4.58^\circ$. (b)-(d) The curl of relaxation displacement vectors in $L_2$ at the global minimum, (b) below, (c) at, and (d) above the bifurcation point.}
\label{fig:bifurcation}
\end{figure*}

Fig.~\ref{fig:bifurcation}a shows the total energy after relaxation as a function of twist angle in twisted trilayer parallel $\rm{WSe_2}$ for the two types of solution, confirming the existence of a critical bifurcation point at $\theta_c \approx 4.58^\circ$. Above this $\theta_c$, the symmetric solution is the only solution. Below $\theta_c$, the asymmetric solution is the global minimum and the symmetric solution becomes an unstable saddle point and has slightly higher energy. Fig.~\ref{fig:bifurcation}b-d shows the three types of local minima below, at, and above the critical twist angle. In particular, Fig.~\ref{fig:bifurcation}c shows that at $\theta_c$, the asymmetric solution has very weak pinwheel-like rotation and it almost becomes a symmetric one. While we use twisted parallel $\mathrm{WSe_2}$ as an example, the bifurcation also occurs in graphene or any trilayers with inversion symmetry between AB and BA stacking orders.

\section{Conclusion}\label{sec:conclusion}
We have presented a general formalism to obtain relaxation patterns in twisted trilayer systems, which cannot be achieved with the traditional continuum model in real space due to the aperiodicity of the system. We give results for twisted trilayer graphene and $\mathrm{WSe_2}$. We introduce configuration space as a natural description of the twisted trilayer system. We show that the relaxation pattern of trilayer systems is the ``moir\'{e} of moir\'{e},'' as a result of the incommensurate coupling between two bilayer moir\'e lengths that correspond to the two twist angles. The relaxation pattern manifests the nonellipticity of the energy minimization problem in the sense that there are infinitely many large length scales emerging in the system, and our model is capable of capturing these large length scale modes. Similar to the bilayer relaxation, in trilayers, the relaxation forms domains at the bilayer moir\'e lengths to maximize the equilibrium stacking orders and minimize the high energy stacking order (AA stack). In contrast to the in-plane relaxation pattern in twisted bilayer graphene where the symmetry of the system is preserved, we found that the twisted trilayers can have symmetry-breaking relaxation patterns. When $\theta_{12} = \theta_{23}$, the trilayer relaxation has two types of solutions: one symmetric solution that preserves the two-fold rotational symmetry along the $z$-axis and the other asymmetric solution that breaks this rotational symmetry. This symmetry breaking might be experimentally observable effect, for example, through second harmonic generation or scanning transmission electron microscopy.

Although we have only presented results for homo-trilayers in this work, our approach can be also applied to hetero-trilayer materials. For materials with similar lattice constants (such as graphene and hBN), we can simply apply a small strain to the monolayer by assuming all three layers  have the same lattice constant. In this way, we can calculate the GSFE by performing rigid shifts in a unit cell using DFT total energies as the interlayer coupling, which effectively takes care of the small strain. For materials with very different lattice constants, for example, graphene on top of transition metal dichalcogenides, we need to modify the definition of $A_i$ matrices and be careful about mapping from configuration space to real space. The configuration space approach can also be generalized to four or more layers, and each additional layer adds two dimensions to configuration space. However, the extension would be very computationally expensive - the computational cost scales as $N^{2(d-1)}$, where $d$ is the dimension and $N$ is the discretization. Therefore, more sophisticated minimization methods should be introduced for computational efficiency. Moreover, we would expect a symmetry-breaking solution to be present for heterostructures with an odd number of layers.

The mechanical relaxation effect is an important correction to be taken into account for in modeling the electronic structure of twisted trilayers. The relaxation pattern in twisted trilayer may be used as a new platform to engineer strain in vdW heterostructures. The second twist angle is an additional degree of freedom to manipulate the electronic states and explore unconventional strongly correlated states~\cite{tsai2019correlated}.

\begin{acknowledgements}
We thank Stephen Carr, Shiang Fang, Steven Torrisi, and Emine Kucukbenli, Philip Kim, Liang Fu, Ke Wang for insightful discussions. This work was supported by the STC Center for Integrated Quantum Materials, NSF Grant No. DMR-1231319, ARO MURI Award W911NF-14-0247, NSF DMREF Award 1922165, and NSF Award DMS-1819220. Calculations were performed on the Odyssey cluster supported by the FAS Division of Science, Research Computing Group at Harvard University.
\end{acknowledgements}

\begin{appendices}
\section{Generalized Stacking Fault Energy}\label{sec:gsfe}
The GSFE is the total energy at different stacking configurations without rotation between a bilayer. Letting $(b_x, b_y)$ be the relative displacement or the local stacking order between two adjacent layers, we define the following vector $\bm v= (v, w) \in [0, 2 \pi ]^2$:
\begin{equation}
	\begin{pmatrix}
		v \\ w
	\end{pmatrix}
		= \frac{2 \pi}{\alpha} \mqty[\sqrt{3}/2 & -1/2 \\ \sqrt{3}/2 & 1/2]
	\begin{pmatrix}
		b_x \\ b_y
	\end{pmatrix}.
\end{equation}
In terms of $\bm v$, the GSFE of a hexagonal lattice can be expressed as a Fourier sum as \cite{kaxiras1993free, zhou2015van, carr2018relaxation},
\begin{align}
V^\mathrm{GSFE}_{j\pm} = c_0 + & c_1(\cos v + \cos w + \cos (v + w) ) \nonumber \\
+ & c_2 (\cos (v + 2w) + \cos(v-w) + \cos(2 v + w)) \nonumber \\
+ & c_3 (\cos(2 v ) + \cos (2 w) + \cos(2 v + 2 w)) \nonumber \\
+ & c_4 (\sin v + \sin w - \sin (v + w))   \label{eqn:vgsfe}\\
+ & c_5 (\sin (2 v+ 2 w) - \sin (2 v) - \sin(2 w)),\nonumber
\end{align}
where we use the subscript $j\pm$ to denote the GSFE between the $L_j$ and $L_{j\pm1}$. We note that we are only including the first few terms in the Fourier series because of the lattice symmetry and higher order terms are three or more orders of magnitude smaller than the leading order terms. For other lattice geometries, it may be necessary to include higher order terms in the Fourier series.

\section{Density Functional Theory Calculations}\label{sec:dft}

The coefficients for the interlayer and intralayer energies are obtained from DFT calculations with the Vienna Ab initio Simulation Package (VASP)~\cite{kresse1993ab, kresse1996efficient}. The van der Waals force is impletemented through the vdW-DFT method using the SCAN+rVV10 functional~\cite{dion2004van, vydrov2010nonlocal, klimevs2011van,peng2016versatile}. The values are presented in Table \ref{tab:values}. The elasticity constants and interlayer coupling coefficients of graphene and $\mathrm{MoS_2}$ are also given in S.Carr {\it et al}\cite{carr2018relaxation}.

To obtain the GSFE, we performed a rigid shift between a bilayer using a $9 \times 9$ grid in the unit cell. We extract the ground state energy from DFT, by allowing relaxation in the out-of-plane direction but fixing the in-plane direction. Fig. \ref{fig:gsfe} shows the GSFE of graphene and $\mathrm{WSe_2}$, with high-symmetry stacking orders, AA stacking and AB/BA stacking, marked. In graphene, the AB/BA stacking (when the A sublattice of one layer is aligned with the B sublattice of the second layer) is the equilibrium stacking configuration and the AA stacking (the origin) has the highest energy. In TMDCs, $\mathrm{MX_2}$, with M being a transition metal atom (e.g., Mo, W) and X being a chalcogen atom (S, Se, Te), the bilayer can have two orientations when performing the rigid shifts because the two sublattices have different atomic species. In parallel $\mathrm{WSe_2}$, the two bilayers are aligned. The GSFE energy landscape is similar to that of graphene. In anti-parallel $\mathrm{WSe_2}$, however, the inversion symmetry is broken. In this case, the equilibrium stacking order is when the metal atoms and the chalcogen atoms are aligned (AB stacking).

The GSFE is a more accurate description of interlayer coupling compared to empirical
interlayer coupling potentials when they exist, such as for graphene and hBN, and it enables the description of interlayer coupling in TMDCs for which empirical interlayer coupling potentials do not exist.

\end{appendices}

\bibliography{mybib}

\end{document}